\documentclass[a4paper,fleqn,usenatbib]{mnras}
\usepackage{graphicx}
\usepackage{amsmath}
\usepackage{amssymb}
\usepackage[T1]{fontenc}
\usepackage{ae,aecompl}
\usepackage{graphicx}
\usepackage{newtxtext,newtxmath}
\usepackage{lscape}
\usepackage{ulem}

\usepackage{xcolor}
\begin{document}
\title[Stopping on the slope]{Halting Migration in Magnetospherically Sculpted Protoplanetary Disks}

\author[Yu, Hansen \& Hasegawa]{Tze Yeung Mathew Yu$^{1}$ \thanks{email:tzeyu@astro.ucla.edu},
Brad Hansen$^{1}$ and Yasuhiro Hasegawa$^2$\\
$^1$Mani L. Bhaumik Institute for Theoretical Physics, Department of Physics and Astronomy, University of California, Los Angeles \\
$^2$Jet Propulsion Laboratory, California Institute of Technology, Pasadena, CA 91109, USA}

\date{accepted:}

\maketitle

\begin{abstract}
We present a physically motivated model for the manner in which a stellar magnetic field sculpts the inner edge of a protoplanetary disk,
and examine the consequence for the migration and stopping of sub-Neptune and super-Earth planets. This model incorporates a transition zone
exterior to the inner truncation of the disk, where the surface density profile is modified by the diffusion of the stellar magnetic field into the disk.
This modification results in a migration trap at the outer edge of the transition zone.
We performed simulations of single planet migration, considering a range of stellar magnetic field strengths and magnetic diffusion profiles. Our simulations show a tight relationship between the final locations of planets and the total magnetic budget available for the disk from their host star. We found that a stellar magnetic field between 67 to 180G and a power-law index between 3 and 2.75 can reasonably reproduce the location at which the observed occurrence rate of  close-in Super-Earth and Sub-Neptune populations changes slope.

\end{abstract}


\section{Introduction}

The discovery of planets in short period orbits, both of the Jovian \citep{MQ95,BM97,Noyes97} and sub-Jovian 
classes \citep{Rivera05,Beau06,UBD07,Charb09,Leger09,Bor11}, suggests that our understanding of planet formation, as based on the Solar system, is significantly incomplete. In particular, the large frequency of occurrence of planets interior to Mercury's orbit illustrates the importance of processes that accumulate planets in the interior regions of protoplanetary disks.

Proposals for how this occurs are varied. Individual planets may be tidally captured at late times, and
circularised into short period orbits \citep{RF96,WM96}, but stable multiple planet systems are difficult to form in this way. In this case, it is generally agreed that material migrates inwards through interaction with the gaseous protoplanetary disk, although there is still debate as to whether this occurs as fully evolved planets
 \citep{LBR96,W97,KN12}, or whether as lower mass entities that assemble later \citep{HM12,CT14}, and which may mimic in situ formation \citep{BHL00,CL13,HM13,BGG16,BBL16}.

An important question in these latter models is what causes the inward migration to halt? The process of inward migration is dependant on the properties (such as temperature and surface density) of the gaseous disk, and a variety of processes can serve to trap planets at special locations (e.g. \cite{KL02,Tsang11,HP11,UOF17,ML18}).
 The most commonly used model assumes that planets migrate to the inner edge of the protostellar disk, which 
 is believed to be truncated by pressure from the stellar magnetic field. A large fraction of models for planet migration invoke a simple truncation of the disk at this edge, either as a true step function or a smoothed transition over narrow radial range \citep{LBR96,RL06,Pap07,TP07,ACL09,CGB10,LC17,IOR17,BMM18}.

The observational evidence regarding the occurrence frequency of planets does support the notion of structure at the inner edge of the planetary disk. Giant planets seem to exhibit a preference for orbital periods in the range 1--5~days, flanked by a sharp drop-off in frequency in both directions. Planets in the sub-Neptune/super-Earth range, on the other hand, show a frequency that rises at short orbital periods, peaking at $\sim 6$--12~days, followed by a slowly decreasing frequency at larger separations \citep{How10,How12,May11,You11,Peti13,Peti18}.

Thus, while a relatively sharp cutoff seems to be qualitatively consistent with the Jovian-class planets, the lower mass population requires a model that distributes the planet frequency more broadly. There have
been various proposals to broaden the influence of the disk inner edge. \citet{LC17} suggest that the disk inner edge is set by the requirement that the disk co-rotate with the stellar magnetic field, and therefore explain the range of inner edge separations with a range of stellar rotation rates. On the other hand, \citet{LOL17} suggest that the magnetospheric cavity will expand with time, as the accretion rate drops and the ram pressure of inflowing material decreases, allowing the magnetic field pressure to push out. The final planet location will then be determined by when the planet `freezes out' of the migration process because the disk density gets too low.

However, these models still use the basic idea of a sharp inner edge to the protoplanetary disk. Yet, the very notion of a simple disk truncation is rather naive. Detailed models do exist for the truncation of gaseous disks due to the action of a central stellar magnetic field, first
 developed for accretion disks around compact objects \citep{GL77,GL79,Kon91}. In particular, the idea that the truncation occurs in a narrow radial range actually violates one of the core principles of those models, which is that a current flow in a narrow annulus (such as the inner edge of an accretion disk) cannot completely screen the stellar magnetic field from the disk. A proper model for the
structure of the inner disk must contain two transition regions -- an inner, narrow one which is effectively the boundary layer where the gas is brought from quasi-Keplerian rotation to co-rotation with the stellar magnetosphere -- and a second, more extended, zone where the stellar magnetic field diffuses into the disk and contributes to the effective viscosity of the gas flow. Our goal is to provide such a model and to examine the consequences for planetary migration.

In \S~\ref{s:DiskModel} we describe a model for the inner structure of a protoplanetary disk that incorporates the two zone model for the interaction of stellar magnetic field with protoplanetary disk, including the treatment of disk evolution and how the magnetospheric cavity evolves. In \S~\ref{S:method_models} we describe how the differences between the disk structure, relative to other models, affects the inward migration of planets of different mass, and in \S~\ref{s:Result_f} and \S~\ref{s:Result_m} we show how this determines the final location of migrating planets.
In \S~\ref{s:discussion}, we compare our results with observations and those of previous studies and discuss some caveats and the future work.
\S~\ref{s:sum_con} is devoted to the summary and conclusion of this work.

\section{The Two-Zone model for Disk--Magnetosphere Interaction}\label{s:DiskModel}

For this work, we considered single planets embedded in a one dimensional viscous disk model.
This disk is assumed to revolve around a Solar mass pre-main sequence star.
We solve for the heating and cooling to provide a self-consistent structure for the temperature and surface density in the radial direction.

\subsection{The truncation of the protoplanetary disk by the stellar magnetic field} 
\label{ss:disk_model}

\begin{figure}
    \centering
    \includegraphics[width=0.8\linewidth]{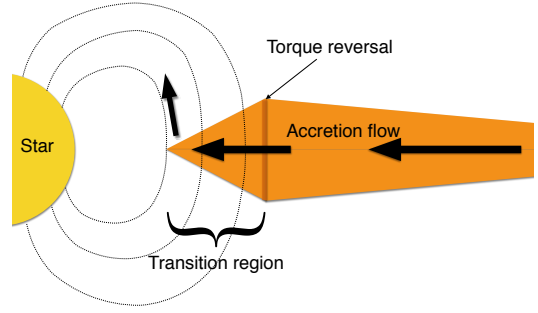}
    \caption{Overview of our disk model. The disk can be separated into 3 regions: the inner truncation, the transition region and the outer disk. The 3 distinct regions are characterized by the amount of stellar magnetic influence they received, from highest to lowest. We
    find that the migration torque experienced by planets can change sign at the outer edge of the transition region.}
    \label{fig:disk_model}
\end{figure}

Our model for how the stellar magnetic field interacts with the protoplanetary disk is based on the work of \cite{GL79}, previously applied to accreting compact objects.
In this model, the disk has 3 distinct regions: the inner truncation, the transition region and the outer disk.
A diagram of the overall disk structure can be found in fig~\ref{fig:disk_model}.
The three regions are determined based on the amount of stellar magnetic influence, with strongest to weakest.

\subsubsection{Inner Truncation}\label{ss:inner_truncation}
First, we consider a magnetic truncation caused by the dipole stellar magnetic field overwhelming the ram pressure of the accreting gas, as in most other models.
If we consider a balance between magnetic pressure and ram pressure at this truncation:

\begin{equation}
    \frac{B_0^2}{8\pi} \left(\frac{R_*}{R_{tc}}\right)^6\approx\frac{1}{2}\rho V^2,
    \label{eq:B_P_balance_app}
\end{equation}
where $R_*$ is the stellar radius, $B_0$ is the magnetic field at $R_*$, $\rho$ is the density of the accretion flow, $R_{tc}$ is the radial distance from the star at which the truncation exists and $V$ is the Keplerian velocity.

Then, if we assume the accretion flux from the disk onto the star is conserved, the location of this truncation is given by

\begin{equation}
\begin{split}
    R_{tc} = & 0.041AU\left(\frac{B_0}{10^3\textit{G}}\right)^{4/7}\left(\frac{M_*}{M_\odot}\right)^{-1/7}\left(\frac{R_*}{1.5R_\odot}\right)^{12/7} \\
    & \left(\frac{\Dot{M}}{10^{-8}M_\odot yr^{-1}}\right)^{-2/7},
\end{split}
\end{equation}\label{eq:truncation_main_text}
where $M_*$ is the mass of the host star, $M_\odot$ is the mass of the Sun, $R_\odot$ is the radius of the Sun, and $\Dot{M}$ is the accretion rate of the disk.
The outcome of this derivation is basically identical to that of \cite{LOL17} (their equation (6)), namely a sharp drop in surface density. The surface density interior to this truncation is assumed to be zero.

\subsubsection{Transition Zone}\label{sss:trans_zone_alpha}

The principal difference between our and prior models is that, exterior to the inner truncation, we include a  transition zone, as described in \cite{GL79}, that is threaded through by the stellar magnetic field.
In this transition zone, the stellar magnetic field has strong influence on the disk gas evolution.
Specifically, the stellar magnetic field increases the effective viscosity ($\alpha$) that the gas experiences as the stellar magnetic field transfers angular momentum to and from the gas.
The strength of this additional viscosity will depend on how easily the magnetic field is able to diffuse into the disk, and therefore the strength of ambipolar diffusion and magnetic field advection.
We will therefore consider a range of magnetic field profiles, parameterised by the power law index $n$ (a value $n=3$ indicates the unmodified stellar dipole).  
 We follow the effective viscosity ($\alpha$)-magnetic field prescription described in \cite{Armitage16} and \cite{SSAB16} which takes the following form:

\begin{equation}
    \alpha = \frac{11}{\beta^{0.53}}; \, \, \, {\rm where} \, \, \,  \beta =\frac{\Sigma/\sqrt{2\pi} H \Omega^2}{B_{loc}^2/8\pi},
    \label{eq:effective_alpha_main}
\end{equation}
and $\Sigma$ is the disk gas surface density, $H$ is the disk vertical scale height, and $\Omega$ is the Keplerian orbital angular frequency.

The magnetic field in the above formulation ($B_{loc}$) is the local magnetic field at a given semi-major axis: 

\begin{equation}
    B_{loc} = B_{nor} \left(\frac{R_{*}}{R}\right)^{n},
    \label{eq:B_loc_main}
\end{equation}
where 
$n$ is the power-law index of the magnetic field profile for the disk mid plane. 
$B_{nor}$ is the normalized magnetic field.
The value of $B_{nor}$ is chosen such that the total magnetic flux passing through the disk between the inner edge  and 2~AU is equal to that of the dipole case ($n=3$) for a given stellar surface magnetic field regardless of their magnetic field profile power-law indices.
The exact equation used for calculating $B_{nor}$ can be found in section~\ref{Ap:mag_diffuse}.
This treatment is done to allow us to compare the influence of the same original stellar field within different models of the diffusivity of the field into the gas, which will yield different values of $n$.

For simplicity, we define the outer radius of the transition zone as a location where the $\alpha$ value drops to $10^{-2}$.
 A discussion of the possible shortcomings of this approach can be found in section~\ref{ss:caveats}.

\subsubsection{Outer Disk}
Beyond the transition zone, the effect of the stellar magnetic field weakens and other sources dominate the viscosity in the disk.
We therefore use a constant $\alpha$ in the outer disk zone, which is set at $10^{-2}$
in this work.

\section{Numerical methods and Model Components}\label{S:method_models}

To investigate how the presence of the transition zone affects the final position of migrating planets, we implement the transition zone
into a standard 1D disk model, which is coupled with an N-body integrator. Here, we describe how each components of the overarching model is implemented along with the numerical setup used for N-body simulations.

\subsection{Host Star}{\label{ss:host_star_model}}

The planetary population observed by the Kepler satellite orbits primarily F, G and K stars.
So, we choose the host star to be a solar mass star in its pre-main sequence stage.
The fiducial parameters we considered are shown in Table~\ref{tb:stellar_para}.
The stellar surface magnetic field is assumed to be dominated by the dipole component.
We vary the stellar surface magnetic field while holding the mass, radius and surface temperature at the value listed.

\begin{table}
   \centering
\begin{tabular}{|l|c|}
    \hline
    \textbf{Parameters} & \textbf{Value} \\
    \hline
    Mass ($M_*$) &  1 $M_\odot$ \\
    \hline
    Radius ($R_*$) & 1.5$R_\odot$ \\
    \hline
    Surface Magnetic Field ($B_0$) & 1000G \\
    \hline
    Surface Temperature ($T_*$ ) & {\color{black}6000K} \\
    \hline
\end{tabular}
\caption{Stellar parameters. $M_\odot$ is the mass of the Sun and $R_\odot$ is the radius of the Sun. We assumed the surface magnetic field is dominated by its dipole component.}\label{tb:stellar_para}
    
\end{table}

 We acknowledge that pre-main sequence stars are known to have drastically contracting radius over a disk life time of a few million years (e.g. \cite{Chabrier97}).
 However, we did not consider the effect of a contracting host star in this project as the contraction should not be substantial while the accretion flow from the protoplanetary disk is still present (\cite{SFB97}).
 Any subsequent contraction after the protoplanetary disk has been dispersed would not affect the planet in the scope of this project.

\subsection{Disk Accretion Rate}{\label{ss:accretion_model}}

\begin{figure}
    \centering
    \includegraphics[width=1\linewidth]{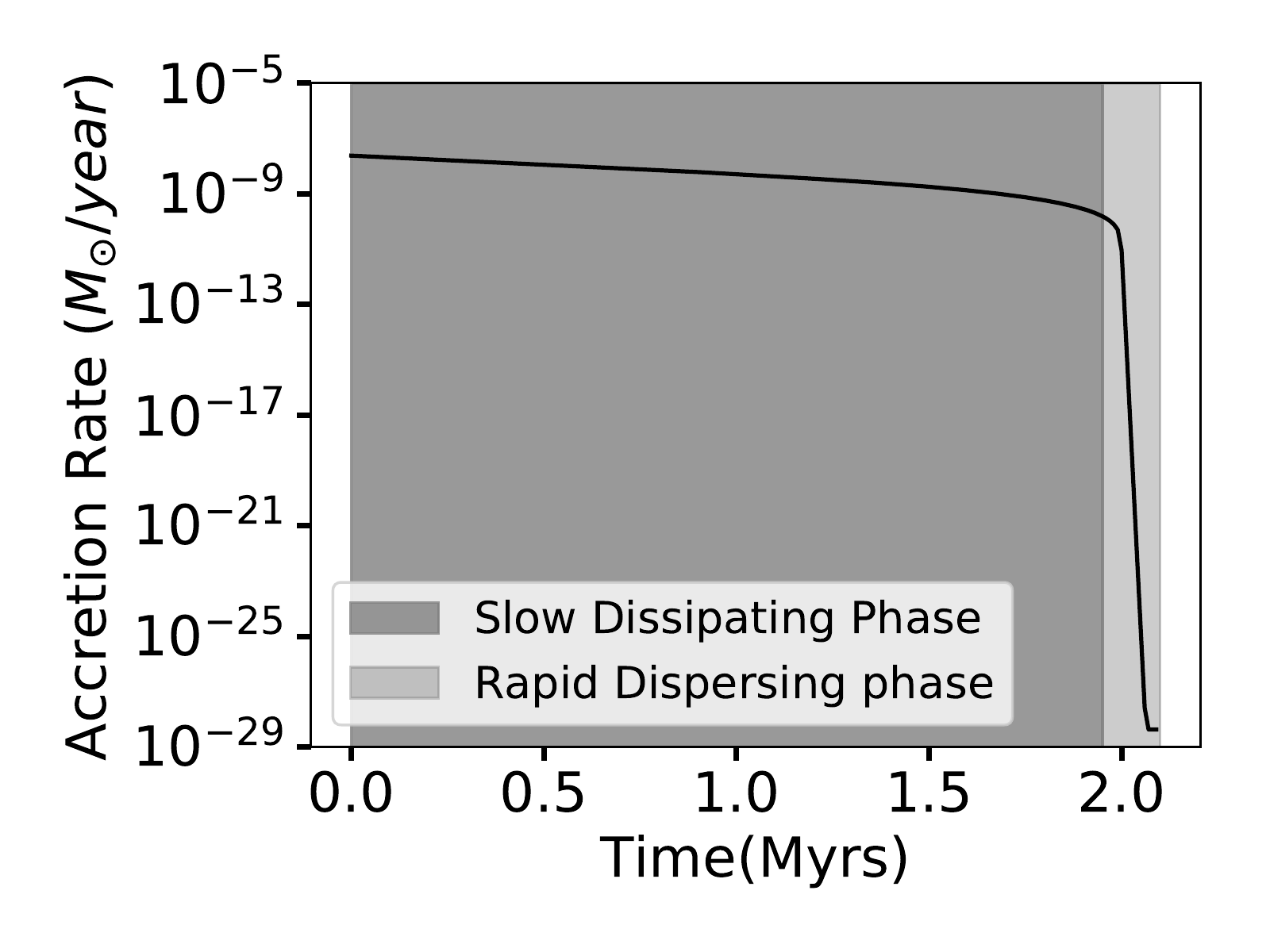}
    \caption{
    The time evolution of the disk accretion rate adopted in our model. 
    The slow dissipating and rapid dispersing phases are shaded by dark and light gray, respectively.
    A slow decrease in accretion rate during the viscous disk phase is followed by a drastic drop. In our model, the drop in accretion rate is assumed to be due to photoevaporation and/ or disk wind opening up a gap exterior to a radius of interest of our model. This gap then shuts off gas and dust supply for the inner disk, causing the accretion rate to drop drastically.
    }\label{fig:accretion_model}
\end{figure}

Planetary migration is determined by the gravitational
interaction between the planet and the natal gas disk, and so planetary migration is also affected by the time evolution of the background protoplanetary disk.
To account for this effect, we compute the accretion rate with time evolution following \cite{AA07}.
This model describes a protoplanetary disk, evolving under the combined influence of viscosity and a  photoevaporative wind off the disk, driven by the high energy radiation from the central star.
The wind eventually opens up a gap at larger radii, cutting off the gas supply to the inner disk, resulting in a rapid drop in the accretion rate as the decoupled inner disk grains onto the star. 
The calculated accretion rate used for this work is shown in Figure~\ref{fig:accretion_model}.
We have shaded the slow dissipating phase and the rapid dispersing phase respectively. The accretion rate initially decreases slowly on a timescale of several Myr, driven by the standard viscous evolution of the disk.
However, at some point ($\sim$2~Myr in the case shown in the figure), the the mass loss rate due to photoevaporative winds becomes comparable to the disk accretion rate, and the winds disconnect the inner disk from the outer disk, shutting off the replenishment of material from larger radii.
After this, the accretion rate onto the star drops precipitously and will have an important influence on when and where the migration will freeze out.

\subsection{Thermal and Density Structures of Disk}\label{ss:Heating_sigma}

To calculate the temperature and surface density profiles, we adopt the steady-state disk model (e.g. \cite{Pringle81}):
\begin{equation}
    \dot{M} = 3 \pi \nu \Sigma,
\end{equation}\label{eq:m_dot}
where $\Sigma$ is the surface density of the disk, $\nu$ is the kinematic viscosity and is assumed to be equal to $\alpha H^2 \Omega$.
No sink other than the central star is considered and the same accretion rate is applicable throughout the disk.

Under the above assumption, we consider the heating of the disk to be caused by both viscous dissipation and passive heating from the host star.
The disk is cooled radiatively.
The passive heating is computed following \cite{UOF17}, which consists of 4 regions.
The regions, described in order of their radial proximity to the host star, are a dust free region, a dust halo region, a dust condensation front, and an opaque thick disk region.
Our temperature profile and surface density profile are then described by the solution to the following equations:

\begin{equation}
    \begin{split}
        \Gamma(R) + \left(\frac{3\Dot{M}}{8\pi}\right)\Omega^2 = \frac{4\sigma T^4}{(3/4\times \kappa\Sigma+1)}
    \end{split}
    \label{eq:heat_cool_equation}
\end{equation}
\begin{equation}
    \begin{split}
    \Sigma =\frac{\Dot{M}}{3\pi}\frac{1}{\alpha H^2 \Omega}
    \label{eq:magnetic_T_body}    
    \end{split}
\end{equation}
where $T$ is the temperature of the disk at the mid-plane. $R$ is the radial location in the disk. $\kappa$ is given in \cite{Bell97}, which formulates their opacity in the form: $\kappa = \kappa_0\rho^aT^b$. $\sigma$ is the Stefan–Boltzmann constant. 
The local passive heating term, $\Gamma (R)$, is defined according to appendix~\ref{apss:passive_heating}.

\subsection{Planetary Migration}\label{s:Migration}

The final element of our model are torques that drive planetary migration.
 As we are mainly interested in Super Earths and Sub-Neptunes, we chose to focus on Type I migration;
 the mass range for Super Earths and Sub-Neptunes should not substantially modify the surface density of the disk for the majority of the disk's lifetime.
 We have checked the applicability of Type I torque for our planets and found that it is valid until less than 0.1~Myrs before the complete dispersion of the disk.
 At that point, the disk surface density is too low to drive any substantial migration.
 
\subsubsection{Type I Migration Prescription}\label{ss:Type1_prescription}
We followed the torque prescription of \cite{Paardekooper11} as implemented in \cite{HN12}, abbreviated as HN12 below.
We included both Lindblad torque (equation~14 of HN12) and corotation effects (equations 15--18, HN12) in our formulation.
Of the corotation torques, we included vortensity-related horseshoe drag, entropy-related horseshoe drag, vortensity-related linear corotation torque and entropy-related linear corotation torque.
We handled the possible saturation of the corotation torque following equation~21 of HN12.
We also included disk induced eccentricity damping and inclination damping following equations~35, 36 and 37.
We applied this formalism, denoted as 2-sided torque below, for the majority of the disk.
The only exception being that we applied a bridging treatment between 1-sided torque and 2-sided torque near the inner truncation of the protoplanetary disk, following the formalism laid out in \cite{LOL17}--equation 9,11 and 14. 
This treatment is needed as the disk is abruptly cut off at the inner edge by definition, creating a step function that the 2-sided torque formalism cannot describe correctly.
We would like to point out that this bridging treatment does not change the overall migration tracks of our simulated planets, with or without the treatment. The inclusion of said treatment is done for completeness sake.

\subsection{Numerical setup}\label{ss:numberical}
We combine all the components discussed in Sections~\ref{s:DiskModel}, \ref{ss:host_star_model}, \ref{ss:accretion_model}, and \ref{ss:Heating_sigma} to develop a time-dependent, 1D disk model that contains the transition zone. The model is then used to compute the torque as described in Section \ref{s:Migration}.
The computed torque is coupled with Mercury N-body simulator \citep{Chambers99} via the user defined force routine, practically. We use the Bulirsch-Stoer algorithm with in Mercury to perform the integration.
 
 To lower computation cost, we pre-generate the disk models before loading them into Mercury. The pre-generated disk models all have radial resolution of $5\times10^{-4}$AU below 1AU and $5\times10^{-3}$AU beyond.
 When values in between radial resolution are required for calculation, they are evaluated via spline fitting.
 The same is performed for the evaluation of the power-law indexes of both temperature and surface density profiles.
 We update the model every 500 years before the simulation time reaches 1.5Myrs and switches to every 50 years to keep up with the faster disk evolution.
 The migration torque is, on the other hand, updated at each time step that Mercury evaluates orbital elements.

For each  set of simulations, we applied the combination of the stellar magnetic field strength and the power-law index listed in table~\ref{tab:Major_sim} to the disk model and performed 100 unique simulations. Each individual simulation contains a single planet and an unique initial location. Planet masses were chosen at integer values between 1 and 10 Earth masses ($M_{\oplus}$), 10 per mass value, and initial semi-major axes were chosen between 0.3 and 1.2 AU in intervals of 0.1AU, 10 per location.
 Initial eccentricities (e) and inclinations (i) are both set to 0.1. All other orbital angles are chosen randomly.

\begin{table*}
    \centering
    \caption{Simulations performed for this project along with their result. Each parameter sets consists of 100 individual simulations. Details of setup can be found  in section~\ref{ss:numberical}. Normalized B-Field shown in this table is a measure of the magnetic budget available for the disk and is calculated assuming an accretion rate of $10^{-8}M_\odot/\textit{years}$. The final location statistic of individual parameter sets performed for this paper are shown in the last 3 columns. Visual representation is shown in figure~\ref{fig:Nor_B-AU}.}\label{tab:Major_sim}
    \begin{tabular}{|c|c|c|c|c|c|c|}
        \hline
        Simulation Name & B-Field Strength (G)  & Power-Law Index(n) & $B_{nor}$ (G) & 1 $\sigma$ lower limit (AU) & median (AU)& 1 $\sigma$ upper limit (AU)\\        
        \hline
         B1000n3 & 1000 & 3.00 & 1000. & 0.2823 & 0.3139 & 0.3395\\ 
         B500n3 & 500 & 3.00 & 500. & 0.2107 & 0.2363 & 0.2502\\ 
         B300n3 & 300 & 3.00 & 300. & 0.1512 & 0.1678 & 0.1834\\ 
         B100n3 & 100 & 3.00 & 100. & 0.0870 & 0.0964 & 0.1037\\ 
         B150n285 & 150 & 2.85 & 116. & 0.0949 & 0.1060 & 0.1155\\
         B300n275 & 300 & 2.75 & 176. & 0.1008 & 0.1439 & 0.1465\\ 
         B150n275 & 150 & 2.75 & 96. & 0.0733 & 0.0803 & 0.0877\\ 
         B100n275 & 100 & 2.75 & 68. & 0.0719 & 0.0738 & 0.0760\\ 
         B100n265 & 100 & 2.65 & 57. & 0.0671 & 0.0709 & 0.0724\\ 
         B100n250 & 100 & 2.50 & 43. & 0.0477 & 0.0481 & 0.0490\\ 
         B100n210 & 100 & 2.10 & 16. & 0.0102 & 0.0108 & 0.0136\\ 
         
        \hline
    \end{tabular}
    
\end{table*}

\section{The effect of the transition zone on planetary migration}\label{s:Result_f}

\label{ss:disk_model_visual}

\begin{figure}
    \centering
    \includegraphics[width=\linewidth]{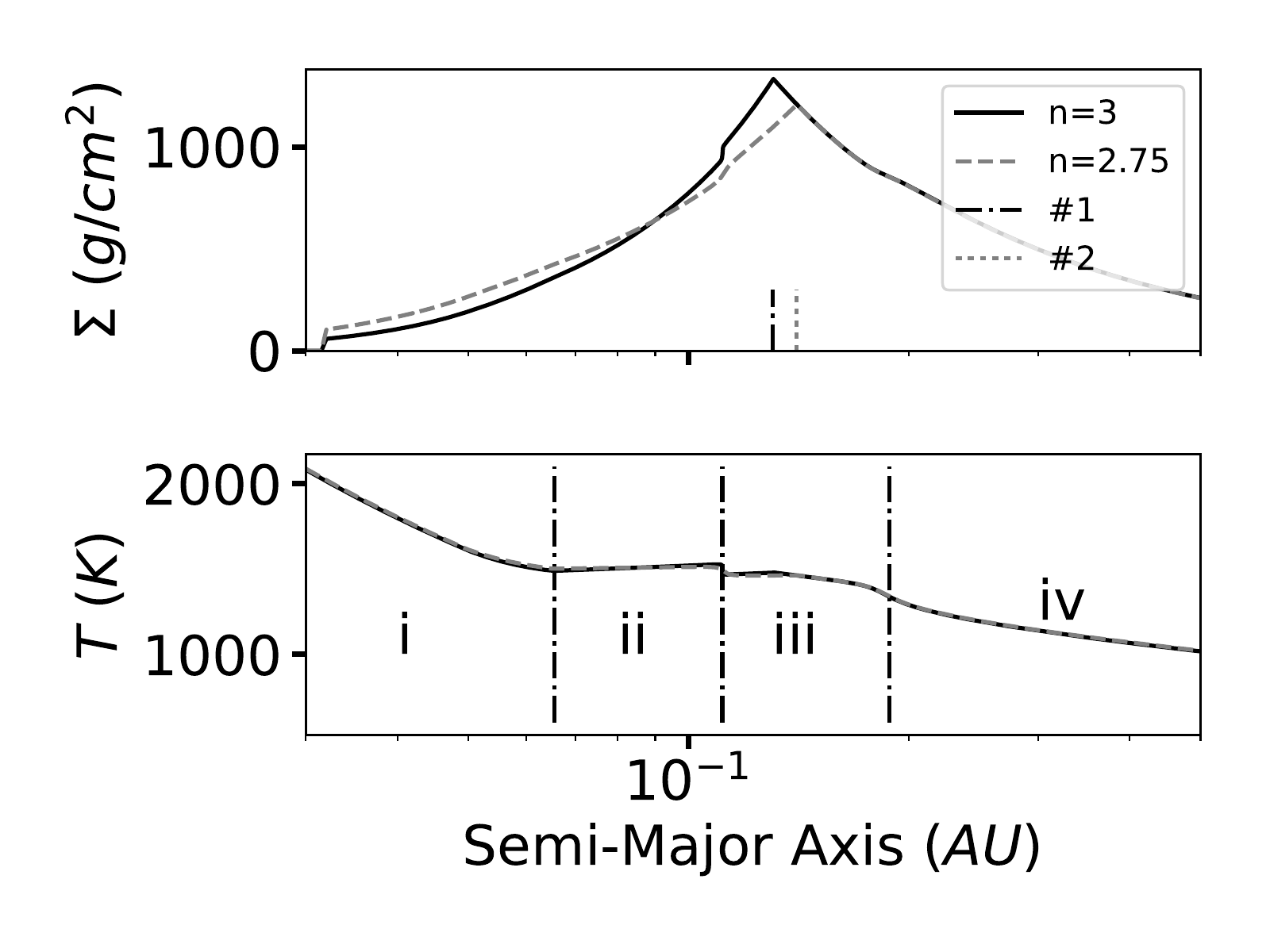}
    \caption{Showing the surface density and temperature profile for $B=1000G$ with power-law index $n=3$ and $n=2.75$. The top panel shows the surface density of the inner 0.5 AU of the protoplanetary disk under the parameter values listed in section~\ref{ss:disk_model_visual}. The bottom panel shows the temperature profile of the same region under the same parameter values. The black solid line denotes the profiles for the fiducial ($n=3$) case and the gray dash line represents the profiles for the $n=2.75$ case. The peaks of the surface density of each surface density profiles are projected onto the radial axis with \#1 and \#2 correspond to those of $n=3$ and $n=2.75$, respectively. On the bottom panel, the two cases show little to no difference in the temperature profile. The temperature zones, described in order of their radial proximity to the host star, are a dust free zone, a dust halo zone, a dust condensation front, and an opaque thick disk zone, which are labeled as i, ii, iii and iv, respectively. On the upper panel, however, the $n=2.75$ case shows a further out peak in surface density and a slower drop off to the host star. The peak locations correspond to the respective planet traps for the two n values. }
    \label{fig:disk_surface_density_temp}
\end{figure}

To place the subsequent discussion in context, here we first describe a snapshot of our complete model for a set of standard parameter values. 
We choose the accretion rate of the disk to be $2\times 10^{-8}M_{\odot}/yr$ and assume that the stellar magnetic field, which has a surface strength of 1000G dominated by its dipole moment ($B\propto$ $R^{-3}$ or $n=3$), going through the disk is unmodified.
The resulting fiducial disk is shown in Figure~\ref{fig:disk_surface_density_temp}.
The surface density and the temperature profile of the disk are shown in the top and bottom panel, respectively.
With this fiducial disk model,
the inner truncation occurs at 0.032AU (approximately 2-day orbit).
The transition zone extends from 0.032AU to 0.138AU (around 2- to 19-days orbit).
As will be described in the next section, the change in surface density slope can lead to reversals in the direction of the migration torque, and so a planet trap results at the outer edge of the transition zone in this snapshot.
Any single migrating planets are therefore expected to be trapped at 0.138AU with a orbital period of 19 days -- farther out than the 2-day location to be expected from a simple magnetospheric truncation model.

The corresponding temperature profile, in the bottom panel of fig~\ref{fig:disk_surface_density_temp}, exhibits similar features to those described in \cite{UOF17}.
The 4 distinct regions as described in \cite{UOF17} are all recognizable with slight modification (see section~\ref{ss:Heating_sigma} for descriptions of regions.).

These regions are labeled as i, ii, iii and iv, respectively. 
The overall temperature is also slightly higher than that of \cite{UOF17} due to our consideration of accretion heating as well as passive irradiation.

\subsection{Migration map}\label{ss:F_migmap}
\begin{figure}
    \centering
    \includegraphics[width=\linewidth]{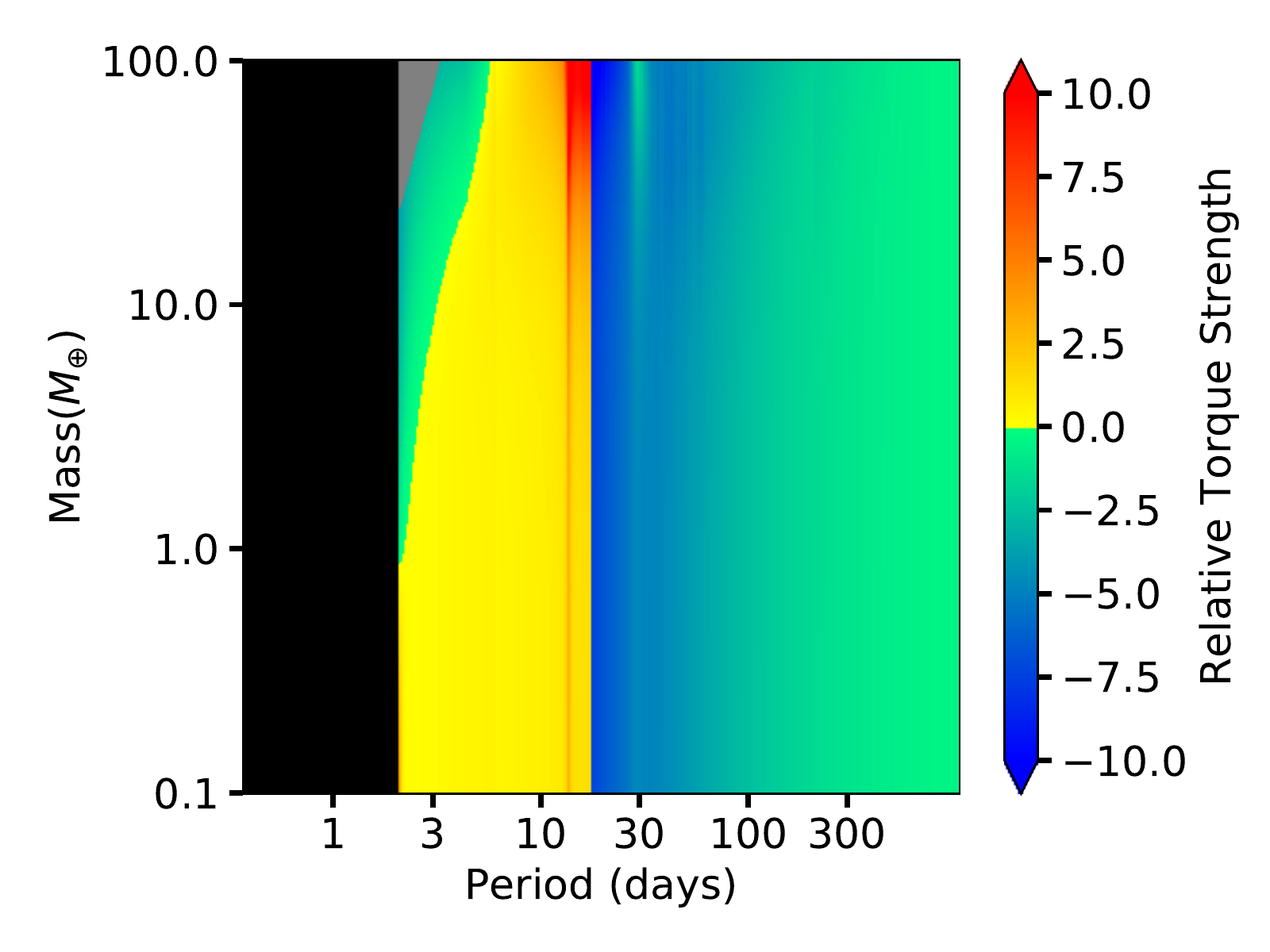}
    \caption{Showing the torque map for $B=1000G$ with power-law index $n=3$. In order to display the features clearly, we normalized the torque values to that of a 1 $M_{\oplus}$ planet at 1AU and then we further removed the mass dependency in the coefficient of the torques by dividing with their respective planet masses squared. A detailed description of the normalization process can be found in appendix~\ref{ap:torque_nor}. The physical torque value for said 1 $M_{\oplus}$ planet is $-1.88\times10^{-17}M_\odot AU^2/\textit{day}^2$ or $-1.12\times10^{33}$ in cgs. Upper and lower limits were applied to the map at 10 and -10 respectively. However, we only observed value $>10$ near 12 days orbit.  The black area on the left denotes the inner cavity where we don't expect migration would be significant due to it's low surface density. The gray triangular area exterior to the cavity at the high mass end denotes parameter space where type II migration become important and cannot be described by our torque formulae. The inner most sliver of green region arises from the 1-sided torque formulation. A torque reversal is observed at around 30 days.}
    \label{fig:disk_map}
\end{figure}

By employing the torque prescription as laid out in section~\ref{ss:Type1_prescription}, we illustrate the nature of the torques for different disk locations and planet masses, using the disk profile of our fiducial model, in Figure~\ref{fig:disk_map}.
The colour map indicates the strength and sign of the calculated torque assuming circular planetary orbits.
The black region indicates that the torque is essentially zero (because the disk surface density is zero inside the magnetospheric cavity 
The colors from green to blue indicate negative torques, which move the planet inwards, towards the star.
Colours from yellow to red indicate positive torques, which will cause the planet to move outwards.
We see that there is a sharp transition from inwardly directed to outwardly directed torques at orbital periods $\sim 30$~days.
This is the location of the peak of the surface density in Figure~\ref{fig:disk_surface_density_temp}, and the boundary between the outer transition zone and the generic viscous disk zone.
This torque reversal implies the existence of a `planet trap', where the migration will stall.

The construction of the torque map is generally governed by the balance between the Lindblad torque and corotation torques.
The blue and green area is dominated by the Lindblad torque while the red and yellow area is dominated by the corotation torques.
The cause of these two distinct regions is related to the value and sign of the corotation torques.
Through out the entire radial extent of the disk, Lindblad torque remains negative.
Even when the power-law index turns from positive to negative moving into the transition zone, both the value and sign of the Lindblad torque remain largely the same.
On the other hand, the corotation torques jump from negative at beyond the transition zone to positive when in the transition zone.
In the transition zone, 
summation of
the corotation torque and the Lindblad torque turns out to be positive and leads to a torque reversal.
It is important to note that the above description applies mainly to low eccentricity cases of $e<0.01$.
As eccentricity increases, the corotation torques weaken and Lindblad torque would once again dominate in the transition region and would effectively erase the torque reversal.
However, in this paper, we only considered single planet cases which, in combination with eccentricity damping from disk interactions, limited the eccentricity to well below 0.01 for the majority of their life time in the disk.

\subsection{Outward Movement of Disk Structure and Lowering in Surface Density} \label{ss:structure_movement}

In the previous section, we consider characteristic disk features at a certain accretion rate.
Here we explore how these features evolve with time, following disk evolution.

When disk accretion rates decreases with time, two major effects related to the disk profile are seen.
First, features of both the surface density and the temperature profile move outward.
Second, the surface density decreases as the accretion rate decreases.
As disk features are crucial in both interpreting the result and comparing with similar works, we have included in Figure~\ref{fig:migration_track1} the time evolution of both the inner edge (black) and the outer edge of the transition zone (orange) throughout the disk's lifetime.
As the accretion rate drops, features in our model tend to move outward.
For the inner edge, the outward movement is due to the magnetic pressure balance requirement.
For the outer edge of the transition zone, the outward movement is due to the effective viscosity ($\alpha$) being inversely proportional to the accretion rate.
So as the accretion rate drops, both of those features moves outward but at different rates and this leads to the overall widening of the transition zone.
On the other hand, the steady decrease in surface density is a consequence of the steady-state assumption (equation~\ref{eq:m_dot}).
We will discuss the effect of these outward-moving features in section~\ref{ss:mig_track} 

\subsection{Planetary Migration, Trapping and Freeze-out}\label{ss:mig_track}

Our disk model represents a refinement of the widespread model for disk-mediated planetary migration by adopting a stopping criterion motivated by a detailed physical model for the manner in which a stellar magnetic field sculpts the inner edge of a protoplanetary disk.
More specifically, we refined the stopping locations of planetary migration in magnetically truncated disks; 
when the transition zone is taken into account, migrating planets can be halted further away from the central star.

\begin{figure}
    \centering
    \includegraphics[width=\linewidth]{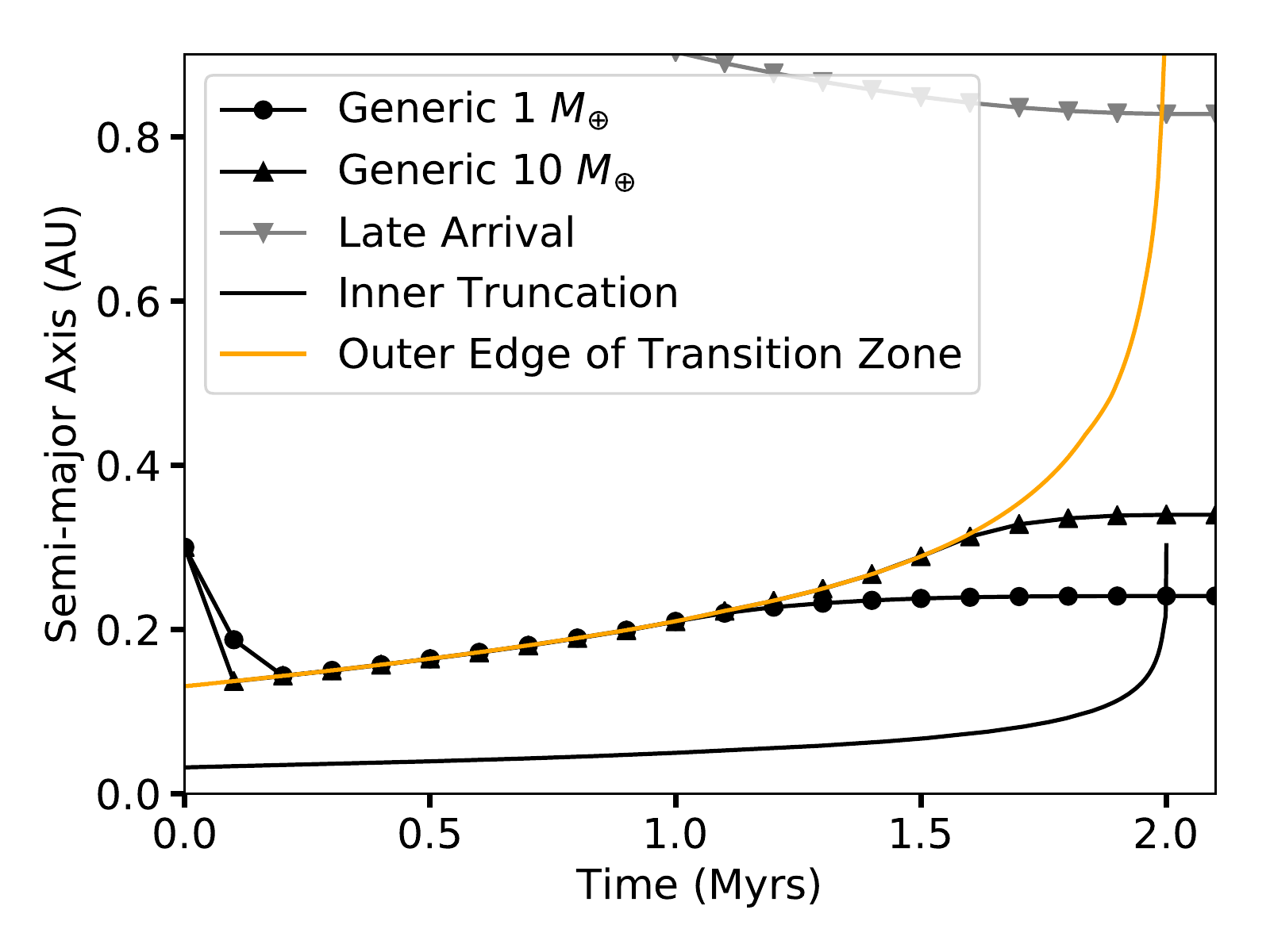}
    \caption{Showing examples of outward movement of the inner truncation and the outer edge of the transition zone, overlaid with the generic and late-arrival migration behavior for the case that $B=1000G$ and $n=3$.
    The inner truncation and the outer edge of the transition zone move outward due to disk evolution (Section~\ref{ss:structure_movement}). For the migration behavior (Section~\ref{ss:mig_track}), in 
    the generic case, planets initially migrate towards the transition zone and become trapped at its outer edge. At later time, the trapped planets decouple from the trap due to weakening of the torques. In the Late-arrival case (Section~\ref{sss:late_trac}), planets migrate inward as well but start at a much larger initial semi-major axis. They do not have enough time to reach the transition zone before the disk is already too dispersed to support much migration. Hence, they are left at larger semi-major axis than the generic counter part.}
    \label{fig:migration_track1}
\end{figure}

To further illustrate the consequences of this effect, we over plotted in Figure~\ref{fig:migration_track1} the typical time-evolutions of the locations of migrating planets for a 1 and 10 $M_{\oplus}$ planet.
We denote them as the generic tracks.

The initial inward migration features planets moving from their starting position towards the planet trap located at the outer edge of the transition zone.
Once planets reach the trap, migration halts and they largely follow the evolution of the disk and remain coupled to the trap.
As the accretion is constantly decreasing in our model, the trap moves outward and brings along any planets that are coupled to it.
At a later time, usually 0.5 to 1 Myrs before the gas disk is completely dispersed, the disk evolution time scale becomes shorter than the outward migration timescales of the planets.
This marks the beginning of the decoupling process, where the planet migrating outward cannot catch up with the outward movement of the outer edge of the transition zone.
The planets now lie interior to the location of the torque reversal and within the transition zone.
So they migrate outward, but at a rate slower than that at which the trapping location and, overall, the transition zone, are moving outwards.
Planets are thus exposed to areas of the disk with ever lower surface density and weaker outward torques. So the planets reach their final location asymptotically.

 The difference in decoupling time between the 1 and 10 $M_{\oplus}$ case is caused by  the planet mass dependence of the torque;
 the 10 $M_{\oplus}$ planet experiences a stronger Type I torque. 
 Due to the stronger torque, higher mass planets are coupled to the planet trap for a longer time and decouple from the disk at more distant locations.
 Further discussion of the effect of planet masses on the final locations of planets can be found in section~\ref{ss:fiducial_loc}.
 The generic track is seen in all $n=3$ simulations, where the planet traps remained similar to the description in section~\ref{ss:F_migmap} through out the lifetime of the disk.

Following the overall picture of the whole migration process, we will discuss in more detail the individual phases and mechanisms of the decoupling process.

\subsubsection{Weakening of Type I Torques and Freeze-out}\label{ss:freeze_out}

As the accretion rate decreases over time in our model, so does the surface density in the disk which directly weakens the torque experienced by the planets.
This change in torque strength leads to two distinct scenarios where planets exit out of their natal disk, which we will discuss more below.

First, planets can completely decouple from the disk evolution and freeze-out at their final location. This case is the typical freeze-out as seen in other works that involve disk migration (e.g.\cite{LOL17}, \cite{IOR17}). The freeze-out in this case refers to when planets cease to migrate and remain at the same semi-major axis for the reminder of and beyond the disk's lifetime. Freeze-out happens when the surface density of the disk can no longer sustain any notable migration. For close to or within the transition zone, freeze-out happens typically around 0.1~Myr before the complete dispersion of the disk.

Second, planets can become decoupled from disk structures they are evolving with. This type of decoupling is most obvious with planets trapped at  the outer edge of the transition zone. It involves the planets' migration time scale becoming longer than the planet trap evolution time scale. In this case, the decoupling does not typically lead to a complete freeze-out immediately as the surface density around the trapping location is still high enough to sustain a limited level of migration. So the planets are still loosely coupled to the disk. The decoupling times vary and are based on the stellar magnetic field strengths and the planet masses. Typically, the decoupling times are earlier for low mass planets compared to high mass planets due to the migration timescales being shorter for higher mass planets. So high mass planets can remain coupled to the disk for longer. The typical decoupling time range from 1~Myr to 0.7Myr before the complete dispersion of our model disk.

\subsection{Late-arriving planets}\label{sss:late_trac}

Although the generic case describes over 90\% of our simulated planets, some planets do exhibit variations in their orbital evolution.
One such variation, which we denote as the late-arrival case, occurs when a planet arrives at the planet trap location at a late time (for instance, if it began migrating from further out).
As the disk is already largely depleted, the strength of the torque reversal at the planet trap location is too weak to hold the planet.
So the planet migrates inwards asymptotically to it's final location, as shown in the grey points in Figure~\ref{fig:migration_track1}.
This late-arrival track is seen in about 5\% of all the simulations.
This type of migration track is typically seen in our simulations when the spawning location of the planet is larger than 0.8AU and with planet masses less than or equal to 2 $M_{\oplus}$. 

\subsection{Final Location Distribution of Planets}\label{ss:fiducial_loc}

\begin{figure}
    \centering
    \includegraphics[width=\linewidth]{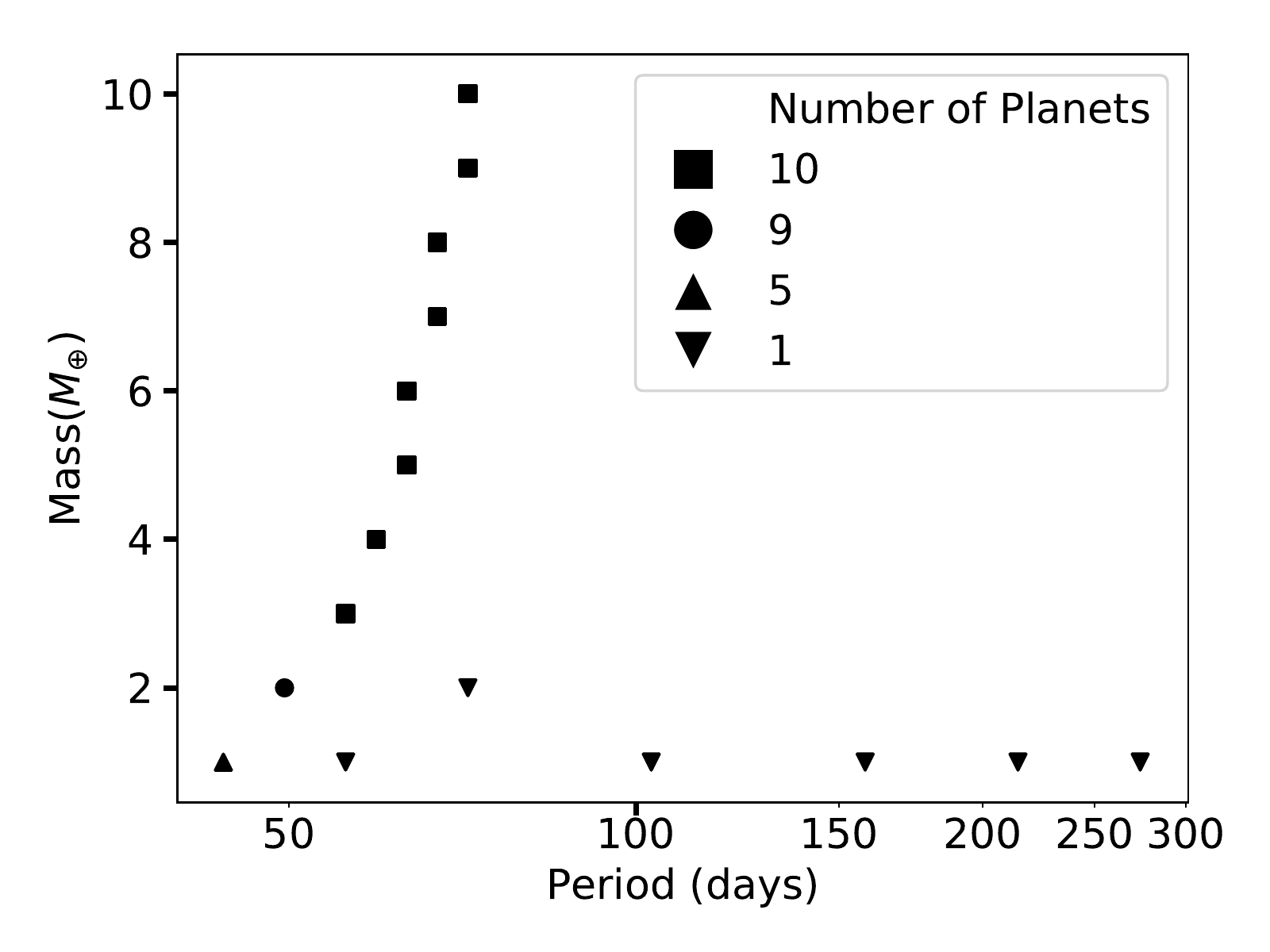}
    \caption{Showing the mass and final orbital period relation for 100 B1000n3 simulations.
    The 100 simulations consist of planets of 10 different planet masses initiated from 10 different locations.
    The different markers show the number of planets at a given final location.
    The majority of the planets that share the same mass also share the same final location, despite different initial positions.
    However, the 1 $M_{\oplus}$ planets spread a much wider range from about 40 days to 300 days.
    The difference in behavior is reflective of the tracks that the planets of different masses reach their final locations. 
    The late arrival track is heavily observed in 1 $M_{\oplus}$ planets simulations.
    On the other hand, the generic track is observed in all but a single 2 $M_{\oplus}$ planet simulation.
    The fundamental difference between the two tracks hinges on whether the planet manages to reach and couple with the planet trap. }
    \label{fig:B1000n3_p-m}
\end{figure}

To illustrate the importance of factors such as planet mass and starting location on the final location, we focus on the simulations set B1000n3, which shares the same magnetic field as our fiducial scenario: $B=1000$G and $n=3$. Figure~\ref{fig:B1000n3_p-m} summarizes the results which show mass and initial location dependency for the lower mass end.

First, we can clearly see that the spread in final location is correlated with planet masses --- with lower mass planets located closer to the host star and the heavier planets located further out.
The only exception to the general trend comes from the stragglers resulted from the late-arrival case and are represented by the 1 and 2 $M_{\oplus}$ planets as seen on the right of the rest of the distribution.

Second, the initial locations of planets have little effect on the final location of the planets except for the least massive planets.
As mentioned in section~\ref{ss:numberical}, a range of initial locations are considered for each set of simulations.
However, despite the difference in initial location, all planets with mass $>3$ $M_{\oplus}$, and the majority of the 2 $M_{\oplus}$ planets, share virtually identical final locations with their peers with the same mass.
The exceptions to this trend are, again, the results of the late-arrival track which is a consequence of the slower migration speeds for lower mass planets.

\section{Result: The effect of magnetic field parameters} \label{s:Result_m}

Our fiducial scenario assumes a dipole magnetic field with strength of 1000G penetrating the disk.
However, depending on the strength of ambipolar diffusion in the disk, the field in the disk may be modified. Furthermore, the surface magnetic fields of stars come in a wide variety and need not be limited to 1000G. We will discuss in this section, first, the effect of changing the magnetic profile inside the disk has on the disk structure, the migration behavior and the final semi-major axis distribution. Then, we will discuss the effect of changing the surface magnetic field on the final location in the context of different magnetic profiles.

\subsection{Disk structure} \label{ss:M_disk_structure}

A high value of magnetic diffusivity in the disk would allow the magnetic field to penetrate further outwards, making the field profile more shallow and lowering the value of $n$.
Alternatively, a low value of diffusivity would make it difficult for the field to penetrate the disk, in the face of the inward advection of material, and therefore this would make the profile steeper, increasing the value of $n$.
Using our normalisation to a fixed global magnetic flux, lowering $n$ would amount to a weaker magnetic field at locations within the disk, while larger $n$ imply stronger local magnetic fields.
A discussion on the exact effect on the magnetic field within the disk can be found in appendix~\ref{Ap:mag_diffuse}.
To demonstrate the effect of changing power law index ($n$) on both the surface density and temperature profiles, we have included the $n=2.75$ case as gray dash lines on fig~\ref{fig:disk_surface_density_temp}.
The main difference between the fiducial case and the $n=2.75$ case is that the peak in surface density moved outward as the power-law index decreased and the drop off in surface density towards the host star is slower in the $n=2.75$ case.
On the other hand, the temperature profile is largely identical.
These behaviors are consistently seen in all non-dipole ($n<3$) simulations. 
The outward shift in peak location and slower drop off rate in surface density have subtle but important effects on the migration process which we will discuss in the next two sections.

\subsection{Migrational Torque and Trapping Location}\label{ss:M_torque}
There are two effects of decreasing n value, namely an increase in radial extension of the transition zone and an overall decrease in torque strength in the vicinity of the outer edge of the transition zone.
For example, the transition zone outer edge extended from 0.13AU to 0.14AU when we decrease from $n=3$ to $n=2.75$ (Figure~\ref{fig:disk_surface_density_temp}).
On the other hand, the torque value experiences a drop of about $5\%$ near the immediate vicinity of the trapping location as a result of a lower surface density and a shallower surface density profile.
The overall consequences of the two competing effects above is a general earlier decoupling time and, hence, smaller final semi-major axis for the planets in the systems.
We will discuss the mechanism that leads to above outcome and the difference in decoupling time between the two example cases in the following section.

\subsection{Effect of Magnetic Field Parameters on Decoupling and Freeze-out}
For the majority of parameters we considered, the behavior of planets are consistent with the fiducial case --- Most planets follow the generic migration track with a small number of them following the late-arrival track. 
This behavior in migration track is true for all magnetic field values we considered.
While the same types of tracks are observed, the decoupling time is observed to be consistently earlier for a given planet mass and given magnetic field strength when n is lowered.
We attribute this earlier decoupling time to the lowering in torque value in the immediate vicinity  of the trapping location as described in section~\ref{ss:M_torque}.
For example, the 1 $M_{\oplus}$ planets in the $n=3$ case decouple at around 1.07Myrs while the same type of planets in the $n=2.75$ case decouple at around 0.225Myrs.
This earlier decoupling time is consistently seen in all simulations with non-dipole ($n<3$) magnetic profile.

Furthermore, we see a deviation from the normal-and-late-arrival track scenario when the power-law index of the magnetic profile (n) drops below 2.65.
The full scope of the behavior seen in that situation can be found in appendix~\ref{AP:decoupling_low_n}

With decoupling time tightly tied to the final location of planets, it is natural to expect an overall smaller final semi-major axis as n decreases.
In the next section, we will discuss the outcome of our simulations with all effect included.

\subsection{Final location of planets}\label{ss:final_loc}

Of the 11 simulation sets,  9 of them, which span $100G \leqslant B  \leqslant1000G$ and $2.5<n \leqslant 3$, behaved similar to our fiducial case (section~\ref{ss:disk_model_visual}).
The median final locations from individual simulation sets range from $7.1\times10^{-2}$AU (7 days) to $3.1\times10^{-1}$AU (64 days), indicating that the model places planets at a wide range of locations, depending on the strength of the magnetic field.
These simulations contain roughly the same mix between generic type and late-arrival type migration tracks. 
This similarity leads them to all have the same mass distribution as described in section~\ref{ss:fiducial_loc}. 
The individual simulation sets statistic, including the $1~\sigma$ equivalences and the median location, can be found in Table~\ref{tab:Major_sim}.

Two sets of simulations yield a behaviour different from the others.
These have power-law index of $n\leqslant 2.5$ and yield median final locations of $<0.05AU$ (4 days).
 The final locations distributions of these two simulations has little to no dependency in mass and are tightly packed at virtually the same location, because the planet trap is either partially deactivated or too weak to halt the migration to the inner truncation of the disk, respectively.
 These cases are discussed in more detail in \S~\ref{apss:dis_trac} and \ref{apss:fall_trac}.

\subsection{Analysis with Normalized B-Field}\label{ss:cn_ana}

\begin{figure}
    \centering
    \includegraphics[width=\linewidth]{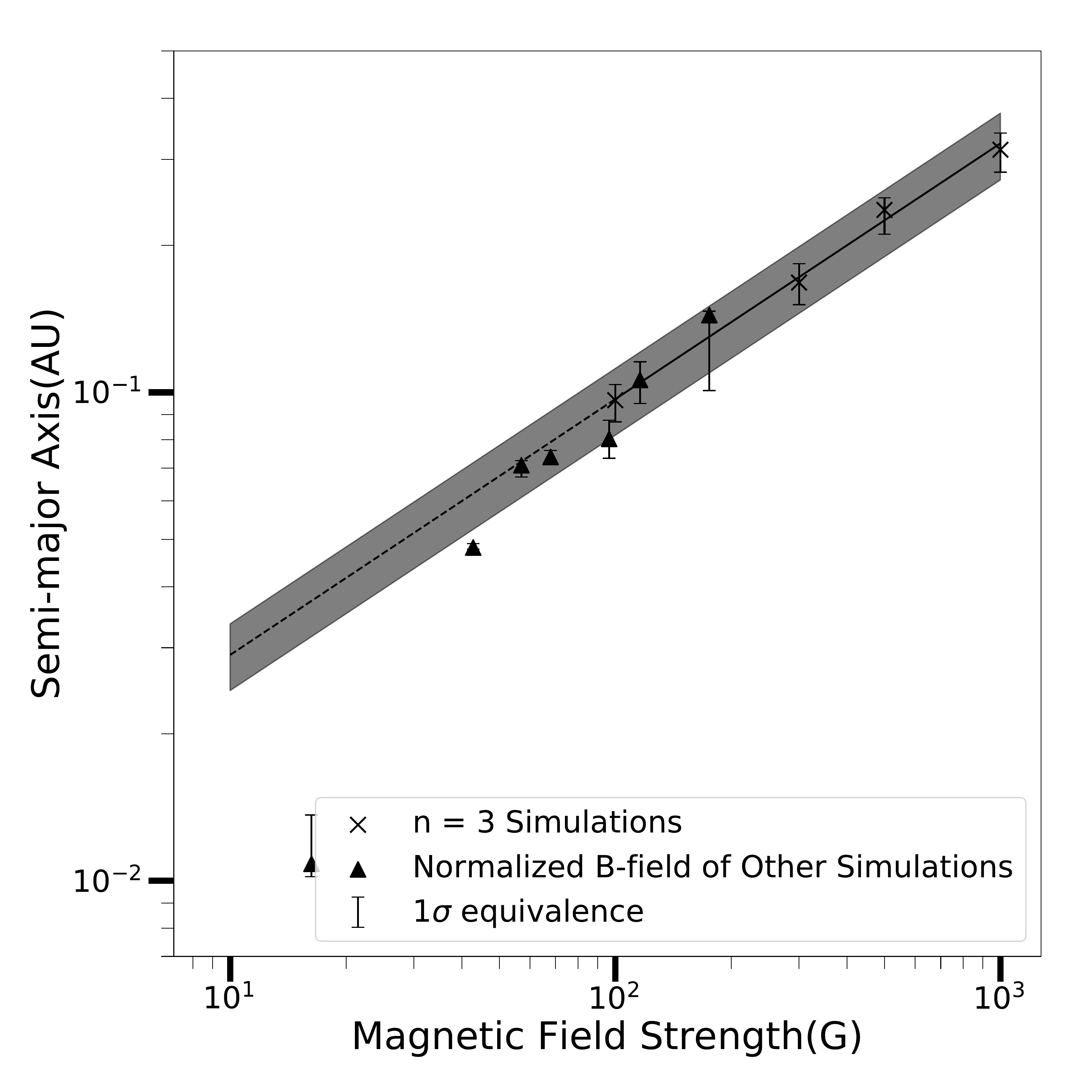}
    \caption{Median Final Locations of each simulation for a given Normalized B-field ($B_{nor}$). Crosses are from the $n=3$ simulations. Solid triangles are from the other simulations. The error bars on the crosses and solid triangles denote the 1-$\sigma$ equivalence of semi-major axis spread in final locations of the same simulation. The solid line, along with the dashed line, shows the best fit relationship along with its extrapolation based on the $n=3$ simulations, respectively. The gray area is the 1-$\sigma$ error of the fit. The fit shows a strong power law relationship between the normalized magnetic field and the final location. While the fit is only based on the $n=3$ simulations, both the fit and its extrapolation appear to represent the relationship between $B_{nor}$ values and the median final locations reasonably well down to approximately 60G. The fitting function can be found in equation~\ref{eq:R-B_fit}.}
    \label{fig:Nor_B-AU}
\end{figure}

Grouping together all simulations we have, we found that the overall result can be described more clearly by designating each of the simulations with their Normalized B-field ($B_{nor}$, see equation~\ref{eq:Cn_not_n2}) instead of their stellar surface B-field ($B_{0}$) or the local magnetic field at the final location of planets.
Disks in simulations with the same $B_{nor}$ are threaded with the same total magnetic flux, even if they have different power laws $n$. 
This correlation may appear counter intuitive, as the normalized B-field is not the one that set the location of the edge of the transition zone -- where the planet trap should reside.
However, if one looks at section~\ref{ss:mig_track}, it is common for planets to have further interactions with the dispersing disk after decoupling begins.
Ultimately, it is perhaps not surprising that the final location correlates more with a global measure of the magnetic influence on this disk than with a localised one.
 In Figure~\ref{fig:Nor_B-AU}, we plotted the final median location against the normalized B-field($B_{nor}$) of each individual simulation set.
 The accretion rate used in the $B_{nor}$ calculation is assumed to be $10^{-8} M_\odot/\textit{year}$. We can see that, when we plot all of our simulation in the normalized B-field and semi-major axis space, the high magnetic field cases, between 1000G and 100G, form a tight relationship in log-log space.
 The solid black fitted function is calculated based on the the $n = 3$ cases only.
 Hence, the fit is independent of the normalization method used.
 Yet, the fit also appears applicable to most of our non-dipole simulations.
 The fit only fails at the lowest field values, when the magnetic effects are not strong enough to halt the migration and the planets migrate to the inner edge.

The fit function is of the following form:

\begin{equation}
R_{final} = (9.7\times10^{-2}\pm 1.4\times10^{-3}AU)\times (\frac{B_{nor}}{100G})^{0.52\pm0.026}
\label{eq:R-B_fit}
\end{equation}

Where $R_{final}$ is the median final location for the given normalized B-field ($B_{nor}$).

Based on the above result, we are confident that the range for which our model can produce a transition zone based planet trap exterior to the inner truncation includes the parameter space $100\leq B_{*} \leq 1000$ with $2.75\leq n \leq 3$.

\section{Discussion}\label{s:discussion}

The simulations we perform demonstrate that the inward migration of Earth and Super-Earth mass planets may be halted before they reach the inner edge of the protoplanetary disk by changes in the disk surface density, mediated by torques exerted by the stellar magnetic field as it diffuses into the disk. In this section, we will compare our result, both the disk structure and the implicated planet distribution, to observation and to other protoplanetary disk models. We will also discuss the caveats and future works.

\subsection{Comparison with Observation}

\begin{figure}
    \centering
    \includegraphics[width=\linewidth]{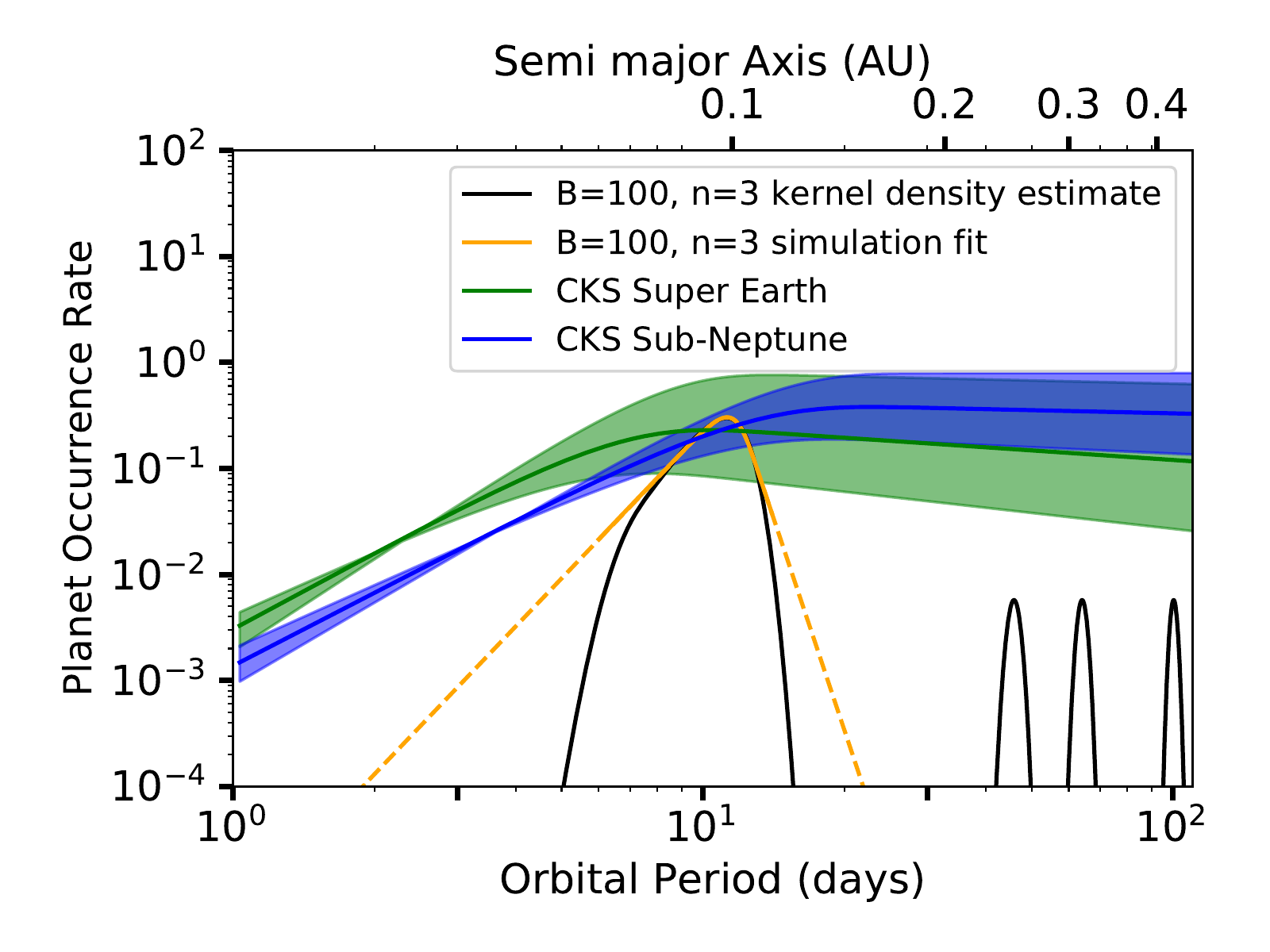}
    \caption{Showing the planet occurrence rates as described in California Kepler Survey(CKS) \protect{\citep{Peti18}} and data from our B100n3 simulations set. The blue and green curves show the occurrence rate for the Sub-Neptune and Super Earth populations, respectively. The shades of the corresponding colors show the errors of the respective occurrence rate. The data from our B100n3 simulation sets is show as both a kernel density estimation and a fit following the convention as described in HOW12. The dash lanes of represent the extrapolation of the fit beyond the radial extent of our data.}
    \label{fig:CKS_compare}
\end{figure}

 We compare the final locations of our simulated planets to observation by directly comparing the result from B100n3 with the Super Earths and Sub Neptune occurrence rate outlined in the California-Kepler Survey IV. (CKSIV ) \citep{Peti18} in Figure~\ref{fig:CKS_compare}.
 The B100n3 simulation is chosen for this comparison as it does not have the complication of normalization as required for the non-dipole profiles while its median final location falls close to the observed drop-off in super Earth occurrence rate.
 One caveat of this comparison is that the CKSIV occurrence rates are calibrated in terms of planet radius, while our simulations are in terms of planet mass, with the relationship to size muddied by unknown envelope fractions. 
 As the mass of Super Earths can be up to $10M_{\oplus}$ and that of Sub-Neptunes can be between $3$ and $20M_{\oplus}$ \citep{Zeng19}, the 1 to $10M_{\oplus}$ mass range in our simulations does not represent a single type of planet but a mix between the two.
 Hence, it is not possible to compare to either one of the population quantitatively. However, a qualitative comparison between the observed occurrence rate and our result can still provide insight into the strengths and shortcomings of our model. Namely, the distribution of the final locations is of most interest.

In Figure~\ref{fig:CKS_compare}, the CKSIV occurrence rates for Super Earth and Sub Neptune are shown as the blue and green curves along with the errors shaded, respectively.
The kernel density estimate of the result of B100n3 is shown as the black curve.
The Gaussian used in the estimate is chosen to have standard deviation of $0.001AU$.
The fit using convention laid out in \cite{How12} (HOW12) is shown in orange.
It is clear that the final location for our simulation is positioned between the drop-offs of Super Earth and Sub Neptune and, hence, can qualitatively match with the drop-off location as seen in the CKSIV fits.
We did not expect an exact match between our result and a specific planet type as we cannot distinguish between Super Earth and Sub Neptune in our simulations.
However, given that our model can place planets in a distribution that mimics the observed population with only single planets in each systems, we consider that our model as promising.

\begin{figure}
    \centering
    \includegraphics[width=\linewidth]{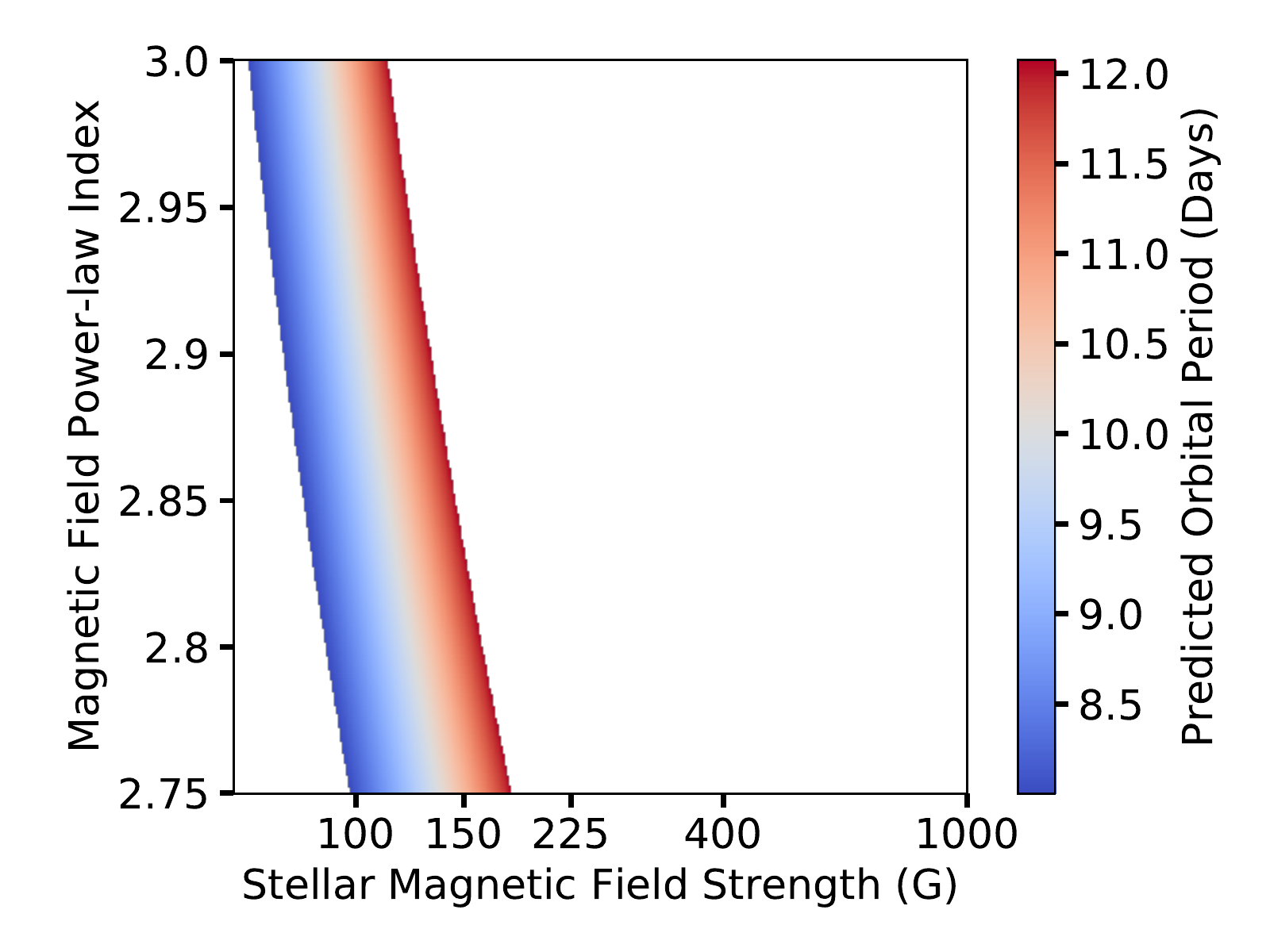}
    \caption{Power-law index (n) and stellar magnetic field strength space that allows planets to freeze-out at the orbital period of 8-12 days. The colored region represents the area of the parameter space that can reproduce the slow drop in occurrence as seen in observations (also see Figure~\ref{fig:CKS_compare}). The color code represents the orbital period at which the median of the planets would be located.}
    \label{fig:n-B_stellar_match_obs}
\end{figure}

Our single planet simulations match the characteristic feature of the turnover in the observed period distribution, but does not match the drop-off below 10 days and the plateau beyond 10 days.
We will address these features in a future publication containing multiple planet simulations.
We postulate that, once more planets are added to our simulation, the plateau area should be filled as interior planets are halted by the planet trap and interact gravitationally with exterior planets.
 The discrepancy of the drop-off at short periods could also potentially be explained by multiple planets as the gravitational influence of later arriving planets may force the innermost closer to the star.
Furthermore, the tidal damping of any eccentricity resulting from dynamical interactions will also drive the innermost planet closer to the star \citep{HM13}. 
Furthermore, using equation~\ref{eq:R-B_fit}, we can then predict the parameter space that can place a planets around the location of the population drop-off.
By restricting the median final location to be between 8 and 12 days orbit (i.e. $7.8\times10^{-2}$ and $1.03\times10^{-1}$AU), we get the applicable normalized B-field values between 67 and 113G.
Mapping these normalized B-field range back into the stellar B-field and stellar power law index space, we get the result in Figure~\ref{fig:n-B_stellar_match_obs}.
The colored region represents the area of the parameter space that planets can be placed in the vicinity where a slow drop in occurrence is observed.
The color code represents the orbital period at which the median of the planets would be located at.
The minimum and maximum stellar magnetic field in this inferred parameter space is 67 and 180G respectively.
As we don't have any dipole based simulation with magnetic field below 100G, we conclude that we are only confident that a stellar magnetic field of 100-180G is able to reproduce observation.
 
Both the full functional range where migrations can be halted by the transition zone (100-1000G) and the observation reproducing range (100-180G) of the magnetic field for our model are supported by observations on T Tauri stars.
As outlined in \cite{JJGDH14}, their observed stellar dipole field strength for T Tauri stars range from 80 to 1720G with a substantial fraction of them being below 400G.
This range matches with the functional range of 100 to 1000G where our model can produce a planet trap associated with the outer edge of the transition zone.
More importantly, a large fraction of the T Tauri star exhibiting dipole field below 400G indicates that our applicable range for matching with the observed super Earth and sub Neptune population is fairly common. 

\subsection{Comparison with protoplanetary disk models}
Our disk model shares sectional similarity with other works in terms of the accretion model considered. On the other hand, we employed a drastically different inner disk from other disk models and that resulted in different migration behavior from other disk models. 

Our accretion model, which is based on \cite{AA07}, can be seen as a bridge between works that consider the long term evolution (e.g. \cite{HG05}, as HG05 below) and those that consider the final dispersion phase of the disk (e.g. \cite{LOL17}, as LOL17 below).

For most of the viscously dominated disk phase, our model behaves comparably with HG05, wherein the
accretion rate is described as a simple power law between the accretion rate and the disk life time, motivated by observations.
The full model in example 1 of HG05 span $10^{-7}$ to $10^{-9}M_{\odot}/\textit{yrs}$ and has form similar to our slow dissipating phase.
This similarity is expected as their model aims at modeling the long term disk behavior which should be dominated by viscous evolution.
However, the migration depends sensitively on how the accretion rate drops towards the end of the disk lifetime, and the drop in $\dot{M}$ in our model is based on a direct calculation of the truncation of accretion due to photoionization of the disk material (see section~\ref{ss:accretion_model}).
This drastic decrease in accretion rate is more abrupt that that used in LOL17, who used an exponential decay model that transitions from a timescale of a few Myr to $10^5$yrs during dispersal.
This is potentially an important distinction, especially for planets that remain coupled with the planet trap to the rapid dispersal phase (i.e. planets with masses higher than 10 $M_{\oplus}$), as it means that the complete freeze-out of migration is even faster in our model.
However, given the upper limit of our mass range, this distinction is not observed in our current result.
On the contrary, we found that the rapid dispersal phase might have minimal effect on planets of mass up to 10 $M_{\oplus}$.
We will discuss the reasoning below based on the decoupling time seen in our simulations.

As most of our planets follow the generic track, planets typically decouple from the disk around 1 to 0.5~Myrs before the complete dispersion of the gas disk.
During this time range, the accretion rate in our model corresponding to $\approx 10^{-9} M_\odot/\textit{yrs}$.
Similar accretion rate occurs in the HG05 model at similar time --- around 0.8~Myrs before the gas disk is assumed to be completely dispersed.
So planets are expected to decouple around that time under the accretion model of HG05.
From the above analysis, we speculate that the final stage of the dispersion process in common models, in combination with our torque treatment, has little to no effect on single planets similar to those in our simulations.

Another important distinction between our model and previous works is that our model for the magnetic field interaction leads to a shallower slope in the surface density in the inner parts of the disk.
Comparing to HG05, which is derived with similar assumption on the heating sources, our temperature and surface density profiles agree largely at the 1AU and beyond scale.
Interior to this, the magnetically enhanced viscosity in our model leads to a surface density is considerably lower with a positive power-law index while the surface density seen in HG05 remains higher than the rest of the disk with a close to 0 or negative power-law index.
This difference in inner disk structures implies that migrating planets are expected to be trapped further out than those in a HG05 disk. Similar argument also applies to other works such as \cite{NGPPSI}, \cite{IOR17} and LOL17 where simple cut-off functions with short scale drop-offs are employed. Hence, our disk model, under the same stellar magnetic field prescription, predicts a further out trapping location compared to other works.

\subsection{Caveats and Possible Future Works}\label{ss:caveats}
While our simulations indicate both a strong mass--final~location trend and normalized-magnetic-field trend, our result requires a few word of caution because of the simplicity of our model on a few aspects, namely, the mass evolution, the lack of multiplicity and viscous $\alpha$ prescription we chose.

First, the mass in our simulation would represent the primordial mass as the planet was still in the disk or right after the disk has been dispersed.
Further mass modification process such as wind driven atmospheric lose or photoevaporation will likely further modify the mass and radius of the planets and lead to possibly a different mass trend that this work does not take into account.
Further work focusing on atmospheric evolution both during and after the disk phase will likely lead to a more realistic mass and size distribution.

Second, we only considered single planets in this work while it is well-known that super Earths frequently come in multiples.
So while our current result points to a rather weak stellar magnetic field, we speculate that the suitable magnetic field range will increase with the introduction of a second or third planet.
This speculation is based on the understanding that planet-planet interaction will likely increase the eccentricity of both planets leading to a weakening of the corotation torques.
So planets would be able to venture deeper into the transition zone.
So, to maintain the median final location at 10-days orbit while still keeping the drop off below 10 days, the transition should occur further out to compensate for this effect.
So the suitable stellar magnetic field value should increase with higher multiplicity.
Future works are planned to explore the effect of multiplicity on the required stellar magnetic field, radial distribution, mass distribution and orbital architecture. 

Third, our viscous $\alpha$ description does not accurately represent the effect from the stellar magnetic field at low magnetic field strength.
As pointed out in section~\ref{sss:trans_zone_alpha}, the expression we used is based on a fit \citep{SSAB16} up to $\beta \lessapprox 10^5$.
However, in order to reach the required $\alpha = 0.01$ for the edge of the transition zone, it requires $\beta \approx 5.5\times10^5$.
While \cite{SSAB16} does not explicitly state that the fit will break down beyond the suggested range, \cite{SMI10}suggest a much shallower relationship should be in place for $10^6 \lessapprox \beta \lessapprox 10^5$.
However, since the effect from the viscous $\alpha$ profile is localized, we expect the modification to our median final locations to be modest.
As the overall effect would require a detailed analysis into the exact balance between the transition zone outward extension and the corotation torque weakening, we leave this topic for future works.

\section{Summary and Conclusion}\label{s:sum_con}

The magnetic influence in the inner part of protoplanetary disks has been considered in other works.
However, the role of stellar magnetic field remained largely limited to producing a truncation at which the accretion flow is balanced by the stellar magnetic pressure.
To improve on that, we constructed a more realistic model by applying a simple transition zone model where the effective viscosity is controlled by the stellar magnetic field.
We found that, within reasonable parameters, said transition zones have the potential of enabling torque reversal which can trap planets radially exterior to the more commonly considered truncation.

By performing single planet simulations using the Mercury N body simulator, we confirmed that trapping can happen for a wide range of parameter space (magnetic field strength (B) and power-law index (n)) within $100G\leq B\leq 1000G$ and $2.75\leq n \leq 3$ and some other cases.
Within a given combination of magnetic field strength and power-law index, we find the final location of lighter planets to be closer to their host star than their heavy counterparts.
Across the range of magnetic field strengths and power-law indexes parameters we considered, we found a power-law relationship between the normalized magnetic fields ($B_{nor}$) and the final median locations of each simulation sets.
This power-law relationship indicates a stronger normalized magnetic field would lead to a larger final median location.
Based on that result, we found that the observed turnover in the low mass planet period distribution can be matched with $67G \leq B \leq 180.0G$ and $2.75 < n \leq 3$.

Overall our model demonstrates that a physically realistic model for the interaction of stellar magnetic field and protoplanetary disk provides a robust explanation for the turnover in the period distribution of the observed occurrence rate of sub-Neptunes and super-Earths. 

\section*{Acknowledgements}
 This research was carried out in part at the Jet Propulsion Laboratory, California Institute of Technology, under a contract with the National Aeronautics and Space Administration (80NM0018D0004) and funded through JPL’s Strategic University Research Partnerships (SURP) program. Y.H. is supported by JPL/Caltech. This research was also supported by NASA grants 80NSSC20K0882 and 80NSSC20K0266. This research has made use of NASA's Astrophysics Data System Bibliographic Services. The High Performance Computing resources used in this investigation were provided by funding from the JPL Information and Technology Solutions Directorate.

\section*{Data Availability}
The data that support the findings of this study are available on request from the corresponding author, T.Y.
 
        \bibliographystyle{mnras}
        \bibliography{main_text}

\begin{thebibliography}{}
\makeatletter
\relax
\def\mn@urlcharsother{\let\do\@makeother \do\$\do\&\do\#\do\^\do\_\do\%\do\~}
\def\mn@doi{\begingroup\mn@urlcharsother \@ifnextchar [ {\mn@doi@}
  {\mn@doi@[]}}
\def\mn@doi@[#1]#2{\def\@tempa{#1}\ifx\@tempa\@empty \href
  {http://dx.doi.org/#2} {doi:#2}\else \href {http://dx.doi.org/#2} {#1}\fi
  \endgroup}
\def\mn@eprint#1#2{\mn@eprint@#1:#2::\@nil}
\def\mn@eprint@arXiv#1{\href {http://arxiv.org/abs/#1} {{\tt arXiv:#1}}}
\def\mn@eprint@dblp#1{\href {http://dblp.uni-trier.de/rec/bibtex/#1.xml}
  {dblp:#1}}
\def\mn@eprint@#1:#2:#3:#4\@nil{\def\@tempa {#1}\def\@tempb {#2}\def\@tempc
  {#3}\ifx \@tempc \@empty \let \@tempc \@tempb \let \@tempb \@tempa \fi \ifx
  \@tempb \@empty \def\@tempb {arXiv}\fi \@ifundefined
  {mn@eprint@\@tempb}{\@tempb:\@tempc}{\expandafter \expandafter \csname
  mn@eprint@\@tempb\endcsname \expandafter{\@tempc}}}

\bibitem[\protect\citeauthoryear{{Adams}, {Cai}  \& {Lizano}}{{Adams}
  et~al.}{2009}]{ACL09}
{Adams} F.~C.,  {Cai} M.~J.,   {Lizano} S.,  2009, \mn@doi [\apjl]
  {10.1088/0004-637X/702/2/L182}, \href
  {https://ui.adsabs.harvard.edu/abs/2009ApJ...702L.182A} {702, L182}

\bibitem[\protect\citeauthoryear{{Alexander} \& {Armitage}}{{Alexander} \&
  {Armitage}}{2007}]{AA07}
{Alexander} R.~D.,  {Armitage} P.~J.,  2007, \mn@doi [\mnras]
  {10.1111/j.1365-2966.2006.11341.x}, \href
  {https://ui.adsabs.harvard.edu/abs/2007MNRAS.375..500A} {375, 500}

\bibitem[\protect\citeauthoryear{{Armitage}}{{Armitage}}{2016}]{Armitage16}
{Armitage} P.~J.,  2016, \mn@doi [\apjl] {10.3847/2041-8213/833/2/L15}, \href
  {https://ui.adsabs.harvard.edu/abs/2016ApJ...833L..15A} {833, L15}

\bibitem[\protect\citeauthoryear{{Batygin}, {Bodenheimer}  \&
  {Laughlin}}{{Batygin} et~al.}{2016}]{BBL16}
{Batygin} K.,  {Bodenheimer} P.~H.,   {Laughlin} G.~P.,  2016, \mn@doi [\apj]
  {10.3847/0004-637X/829/2/114}, \href
  {https://ui.adsabs.harvard.edu/abs/2016ApJ...829..114B} {829, 114}

\bibitem[\protect\citeauthoryear{{Beaulieu} et~al.,}{{Beaulieu}
  et~al.}{2006}]{Beau06}
{Beaulieu} J.~P.,  et~al., 2006, \mn@doi [\nat] {10.1038/nature04441}, \href
  {https://ui.adsabs.harvard.edu/abs/2006Natur.439..437B} {439, 437}

\bibitem[\protect\citeauthoryear{{Bell}, {Cassen}, {Klahr}  \&
  {Henning}}{{Bell} et~al.}{1997}]{Bell97}
{Bell} K.~R.,  {Cassen} P.~M.,  {Klahr} H.~H.,   {Henning} T.,  1997, \mn@doi
  [\apj] {10.1086/304514}, \href
  {https://ui.adsabs.harvard.edu/abs/1997ApJ...486..372B} {486, 372}

\bibitem[\protect\citeauthoryear{{Bodenheimer}, {Hubickyj}  \&
  {Lissauer}}{{Bodenheimer} et~al.}{2000}]{BHL00}
{Bodenheimer} P.,  {Hubickyj} O.,   {Lissauer} J.~J.,  2000, \mn@doi [\icarus]
  {10.1006/icar.1999.6246}, \href
  {https://ui.adsabs.harvard.edu/abs/2000Icar..143....2B} {143, 2}

\bibitem[\protect\citeauthoryear{{Boley}, {Granados Contreras}  \&
  {Gladman}}{{Boley} et~al.}{2016}]{BGG16}
{Boley} A.~C.,  {Granados Contreras} A.~P.,   {Gladman} B.,  2016, \mn@doi
  [\apjl] {10.3847/2041-8205/817/2/L17}, \href
  {https://ui.adsabs.harvard.edu/abs/2016ApJ...817L..17B} {817, L17}

\bibitem[\protect\citeauthoryear{{Borucki} et~al.,}{{Borucki}
  et~al.}{2011}]{Bor11}
{Borucki} W.~J.,  et~al., 2011, \mn@doi [\apj] {10.1088/0004-637X/736/1/19},
  \href {https://ui.adsabs.harvard.edu/abs/2011ApJ...736...19B} {736, 19}

\bibitem[\protect\citeauthoryear{{Brasser}, {Matsumura}, {Muto}  \&
  {Ida}}{{Brasser} et~al.}{2018}]{BMM18}
{Brasser} R.,  {Matsumura} S.,  {Muto} T.,   {Ida} S.,  2018, \mn@doi [\apjl]
  {10.3847/2041-8213/aada18}, \href
  {https://ui.adsabs.harvard.edu/abs/2018ApJ...864L...8B} {864, L8}

\bibitem[\protect\citeauthoryear{{Butler}, {Marcy}, {Williams}, {Hauser}  \&
  {Shirts}}{{Butler} et~al.}{1997}]{BM97}
{Butler} R.~P.,  {Marcy} G.~W.,  {Williams} E.,  {Hauser} H.,   {Shirts} P.,
  1997, \mn@doi [\apjl] {10.1086/310444}, \href
  {https://ui.adsabs.harvard.edu/abs/1997ApJ...474L.115B} {474, L115}

\bibitem[\protect\citeauthoryear{{Chabrier} \& {Baraffe}}{{Chabrier} \&
  {Baraffe}}{1997}]{Chabrier97}
{Chabrier} G.,  {Baraffe} I.,  1997, \aap, \href
  {https://ui.adsabs.harvard.edu/abs/1997A&A...327.1039C} {327, 1039}

\bibitem[\protect\citeauthoryear{{Chambers}}{{Chambers}}{1999}]{Chambers99}
{Chambers} J.~E.,  1999, \mn@doi [\mnras] {10.1046/j.1365-8711.1999.02379.x},
  \href {https://ui.adsabs.harvard.edu/abs/1999MNRAS.304..793C} {304, 793}

\bibitem[\protect\citeauthoryear{{Chang}, {Gu}  \& {Bodenheimer}}{{Chang}
  et~al.}{2010}]{CGB10}
{Chang} S.~H.,  {Gu} P.~G.,   {Bodenheimer} P.~H.,  2010, \mn@doi [\apj]
  {10.1088/0004-637X/708/2/1692}, \href
  {https://ui.adsabs.harvard.edu/abs/2010ApJ...708.1692C} {708, 1692}

\bibitem[\protect\citeauthoryear{{Charbonneau} et~al.,}{{Charbonneau}
  et~al.}{2009}]{Charb09}
{Charbonneau} D.,  et~al., 2009, \mn@doi [\nat] {10.1038/nature08679}, \href
  {https://ui.adsabs.harvard.edu/abs/2009Natur.462..891C} {462, 891}

\bibitem[\protect\citeauthoryear{{Chatterjee} \& {Tan}}{{Chatterjee} \&
  {Tan}}{2014}]{CT14}
{Chatterjee} S.,  {Tan} J.~C.,  2014, \mn@doi [\apj]
  {10.1088/0004-637X/780/1/53}, \href
  {https://ui.adsabs.harvard.edu/abs/2014ApJ...780...53C} {780, 53}

\bibitem[\protect\citeauthoryear{{Chiang} \& {Laughlin}}{{Chiang} \&
  {Laughlin}}{2013}]{CL13}
{Chiang} E.,  {Laughlin} G.,  2013, \mn@doi [\mnras] {10.1093/mnras/stt424},
  \href {https://ui.adsabs.harvard.edu/abs/2013MNRAS.431.3444C} {431, 3444}

\bibitem[\protect\citeauthoryear{{Emsenhuber}, {Mordasini}, {Burn}, {Alibert},
  {Benz}  \& {Asphaug}}{{Emsenhuber} et~al.}{2021}]{NGPPSI}
{Emsenhuber} A.,  {Mordasini} C.,  {Burn} R.,  {Alibert} Y.,  {Benz} W.,
  {Asphaug} E.,  2021, \mn@doi [\aap] {10.1051/0004-6361/202038553}, \href
  {https://ui.adsabs.harvard.edu/abs/2021A&A...656A..69E} {656, A69}

\bibitem[\protect\citeauthoryear{{Ghosh} \& {Lamb}}{{Ghosh} \&
  {Lamb}}{1979}]{GL79}
{Ghosh} P.,  {Lamb} F.~K.,  1979, \mn@doi [\apj] {10.1086/157285}, \href
  {https://ui.adsabs.harvard.edu/abs/1979ApJ...232..259G} {232, 259}

\bibitem[\protect\citeauthoryear{{Ghosh}, {Lamb}  \& {Pethick}}{{Ghosh}
  et~al.}{1977}]{GL77}
{Ghosh} P.,  {Lamb} F.~K.,   {Pethick} C.~J.,  1977, \mn@doi [\apj]
  {10.1086/155606}, \href
  {https://ui.adsabs.harvard.edu/abs/1977ApJ...217..578G} {217, 578}

\bibitem[\protect\citeauthoryear{{Hansen} \& {Murray}}{{Hansen} \&
  {Murray}}{2012}]{HM12}
{Hansen} B. M.~S.,  {Murray} N.,  2012, \mn@doi [\apj]
  {10.1088/0004-637X/751/2/158}, \href
  {https://ui.adsabs.harvard.edu/abs/2012ApJ...751..158H} {751, 158}

\bibitem[\protect\citeauthoryear{{Hansen} \& {Murray}}{{Hansen} \&
  {Murray}}{2013}]{HM13}
{Hansen} B. M.~S.,  {Murray} N.,  2013, \mn@doi [\apj]
  {10.1088/0004-637X/775/1/53}, \href
  {https://ui.adsabs.harvard.edu/abs/2013ApJ...775...53H} {775, 53}

\bibitem[\protect\citeauthoryear{{Hasegawa} \& {Pudritz}}{{Hasegawa} \&
  {Pudritz}}{2011}]{HP11}
{Hasegawa} Y.,  {Pudritz} R.~E.,  2011, \mn@doi [\mnras]
  {10.1111/j.1365-2966.2011.19338.x}, \href
  {https://ui.adsabs.harvard.edu/abs/2011MNRAS.417.1236H} {417, 1236}

\bibitem[\protect\citeauthoryear{{Hellary} \& {Nelson}}{{Hellary} \&
  {Nelson}}{2012}]{HN12}
{Hellary} P.,  {Nelson} R.~P.,  2012, \mn@doi [\mnras]
  {10.1111/j.1365-2966.2011.19815.x}, \href
  {https://ui.adsabs.harvard.edu/abs/2012MNRAS.419.2737H} {419, 2737}

\bibitem[\protect\citeauthoryear{{Howard} et~al.,}{{Howard}
  et~al.}{2010}]{How10}
{Howard} A.~W.,  et~al., 2010, \mn@doi [Science] {10.1126/science.1194854},
  \href {https://ui.adsabs.harvard.edu/abs/2010Sci...330..653H} {330, 653}

\bibitem[\protect\citeauthoryear{{Howard} et~al.,}{{Howard}
  et~al.}{2012}]{How12}
{Howard} A.~W.,  et~al., 2012, \mn@doi [\apjs] {10.1088/0067-0049/201/2/15},
  \href {https://ui.adsabs.harvard.edu/abs/2012ApJS..201...15H} {201, 15}

\bibitem[\protect\citeauthoryear{{Hueso} \& {Guillot}}{{Hueso} \&
  {Guillot}}{2005}]{HG05}
{Hueso} R.,  {Guillot} T.,  2005, \mn@doi [\aap] {10.1051/0004-6361:20041905},
  \href {https://ui.adsabs.harvard.edu/abs/2005A&A...442..703H} {442, 703}

\bibitem[\protect\citeauthoryear{{Izidoro}, {Ogihara}, {Raymond}, {Morbidelli},
  {Pierens}, {Bitsch}, {Cossou}  \& {Hersant}}{{Izidoro} et~al.}{2017}]{IOR17}
{Izidoro} A.,  {Ogihara} M.,  {Raymond} S.~N.,  {Morbidelli} A.,  {Pierens} A.,
   {Bitsch} B.,  {Cossou} C.,   {Hersant} F.,  2017, \mn@doi [\mnras]
  {10.1093/mnras/stx1232}, \href
  {https://ui.adsabs.harvard.edu/abs/2017MNRAS.470.1750I} {470, 1750}

\bibitem[\protect\citeauthoryear{{Johnstone}, {Jardine}, {Gregory}, {Donati}
  \& {Hussain}}{{Johnstone} et~al.}{2014}]{JJGDH14}
{Johnstone} C.~P.,  {Jardine} M.,  {Gregory} S.~G.,  {Donati} J.~F.,
  {Hussain} G.,  2014, \mn@doi [\mnras] {10.1093/mnras/stt2107}, \href
  {https://ui.adsabs.harvard.edu/abs/2014MNRAS.437.3202J} {437, 3202}

\bibitem[\protect\citeauthoryear{{Kley} \& {Nelson}}{{Kley} \&
  {Nelson}}{2012}]{KN12}
{Kley} W.,  {Nelson} R.~P.,  2012, \mn@doi [\araa]
  {10.1146/annurev-astro-081811-125523}, \href
  {https://ui.adsabs.harvard.edu/abs/2012ARA&A..50..211K} {50, 211}

\bibitem[\protect\citeauthoryear{{Koenigl}}{{Koenigl}}{1991}]{Kon91}
{Koenigl} A.,  1991, \mn@doi [\apjl] {10.1086/185972}, \href
  {https://ui.adsabs.harvard.edu/abs/1991ApJ...370L..39K} {370, L39}

\bibitem[\protect\citeauthoryear{{Kuchner} \& {Lecar}}{{Kuchner} \&
  {Lecar}}{2002}]{KL02}
{Kuchner} M.~J.,  {Lecar} M.,  2002, \mn@doi [\apjl] {10.1086/342370}, \href
  {https://ui.adsabs.harvard.edu/abs/2002ApJ...574L..87K} {574, L87}

\bibitem[\protect\citeauthoryear{{Lee} \& {Chiang}}{{Lee} \&
  {Chiang}}{2017}]{LC17}
{Lee} E.~J.,  {Chiang} E.,  2017, \mn@doi [\apj] {10.3847/1538-4357/aa6fb3},
  \href {https://ui.adsabs.harvard.edu/abs/2017ApJ...842...40L} {842, 40}

\bibitem[\protect\citeauthoryear{{L{\'e}ger} et~al.,}{{L{\'e}ger}
  et~al.}{2009}]{Leger09}
{L{\'e}ger} A.,  et~al., 2009, \mn@doi [\aap] {10.1051/0004-6361/200911933},
  \href {https://ui.adsabs.harvard.edu/abs/2009A&A...506..287L} {506, 287}

\bibitem[\protect\citeauthoryear{{Lin}, {Bodenheimer}  \& {Richardson}}{{Lin}
  et~al.}{1996}]{LBR96}
{Lin} D.~N.~C.,  {Bodenheimer} P.,   {Richardson} D.~C.,  1996, \mn@doi [\nat]
  {10.1038/380606a0}, \href
  {https://ui.adsabs.harvard.edu/abs/1996Natur.380..606L} {380, 606}

\bibitem[\protect\citeauthoryear{{Liu}, {Ormel}  \& {Lin}}{{Liu}
  et~al.}{2017}]{LOL17}
{Liu} B.,  {Ormel} C.~W.,   {Lin} D. N.~C.,  2017, \mn@doi [\aap]
  {10.1051/0004-6361/201630017}, \href
  {https://ui.adsabs.harvard.edu/abs/2017A&A...601A..15L} {601, A15}

\bibitem[\protect\citeauthoryear{{Mayor} \& {Queloz}}{{Mayor} \&
  {Queloz}}{1995}]{MQ95}
{Mayor} M.,  {Queloz} D.,  1995, \mn@doi [\nat] {10.1038/378355a0}, \href
  {https://ui.adsabs.harvard.edu/abs/1995Natur.378..355M} {378, 355}

\bibitem[\protect\citeauthoryear{{Mayor} et~al.,}{{Mayor} et~al.}{2011}]{May11}
{Mayor} M.,  et~al., 2011, arXiv e-prints, \href
  {https://ui.adsabs.harvard.edu/abs/2011arXiv1109.2497M} {p. arXiv:1109.2497}

\bibitem[\protect\citeauthoryear{{Miranda} \& {Lai}}{{Miranda} \&
  {Lai}}{2018}]{ML18}
{Miranda} R.,  {Lai} D.,  2018, \mn@doi [\mnras] {10.1093/mnras/stx2706}, \href
  {https://ui.adsabs.harvard.edu/abs/2018MNRAS.473.5267M} {473, 5267}

\bibitem[\protect\citeauthoryear{{Noyes}, {Jha}, {Korzennik}, {Krockenberger},
  {Nisenson}, {Brown}, {Kennelly}  \& {Horner}}{{Noyes} et~al.}{1997}]{Noyes97}
{Noyes} R.~W.,  {Jha} S.,  {Korzennik} S.~G.,  {Krockenberger} M.,  {Nisenson}
  P.,  {Brown} T.~M.,  {Kennelly} E.~J.,   {Horner} S.~D.,  1997, \mn@doi
  [\apjl] {10.1086/310754}, \href
  {https://ui.adsabs.harvard.edu/abs/1997ApJ...483L.111N} {483, L111}

\bibitem[\protect\citeauthoryear{{Paardekooper}, {Baruteau}  \&
  {Kley}}{{Paardekooper} et~al.}{2011}]{Paardekooper11}
{Paardekooper} S.~J.,  {Baruteau} C.,   {Kley} W.,  2011, \mn@doi [\mnras]
  {10.1111/j.1365-2966.2010.17442.x}, \href
  {https://ui.adsabs.harvard.edu/abs/2011MNRAS.410..293P} {410, 293}

\bibitem[\protect\citeauthoryear{{Papaloizou}}{{Papaloizou}}{2007}]{Pap07}
{Papaloizou} J.~C.~B.,  2007, \mn@doi [\aap] {10.1051/0004-6361:20065414},
  \href {https://ui.adsabs.harvard.edu/abs/2007A&A...463..775P} {463, 775}

\bibitem[\protect\citeauthoryear{{Petigura}, {Howard}  \& {Marcy}}{{Petigura}
  et~al.}{2013}]{Peti13}
{Petigura} E.~A.,  {Howard} A.~W.,   {Marcy} G.~W.,  2013, \mn@doi [Proceedings
  of the National Academy of Science] {10.1073/pnas.1319909110}, \href
  {https://ui.adsabs.harvard.edu/abs/2013PNAS..11019273P} {110, 19273}

\bibitem[\protect\citeauthoryear{{Petigura} et~al.,}{{Petigura}
  et~al.}{2018}]{Peti18}
{Petigura} E.~A.,  et~al., 2018, \mn@doi [\aj] {10.3847/1538-3881/aaa54c},
  \href {https://ui.adsabs.harvard.edu/abs/2018AJ....155...89P} {155, 89}

\bibitem[\protect\citeauthoryear{{Pringle}}{{Pringle}}{1981}]{Pringle81}
{Pringle} J.~E.,  1981, \mn@doi [\araa] {10.1146/annurev.aa.19.090181.001033},
  \href {https://ui.adsabs.harvard.edu/abs/1981ARA&A..19..137P} {19, 137}

\bibitem[\protect\citeauthoryear{{Rasio} \& {Ford}}{{Rasio} \&
  {Ford}}{1996}]{RF96}
{Rasio} F.~A.,  {Ford} E.~B.,  1996, \mn@doi [Science]
  {10.1126/science.274.5289.954}, \href
  {https://ui.adsabs.harvard.edu/abs/1996Sci...274..954R} {274, 954}

\bibitem[\protect\citeauthoryear{{Rivera} et~al.,}{{Rivera}
  et~al.}{2005}]{Rivera05}
{Rivera} E.~J.,  et~al., 2005, \mn@doi [\apj] {10.1086/491669}, \href
  {https://ui.adsabs.harvard.edu/abs/2005ApJ...634..625R} {634, 625}

\bibitem[\protect\citeauthoryear{{Romanova} \& {Lovelace}}{{Romanova} \&
  {Lovelace}}{2006}]{RL06}
{Romanova} M.~M.,  {Lovelace} R.~V.~E.,  2006, \mn@doi [\apjl]
  {10.1086/505967}, \href
  {https://ui.adsabs.harvard.edu/abs/2006ApJ...645L..73R} {645, L73}

\bibitem[\protect\citeauthoryear{{Salvesen}, {Simon}, {Armitage}  \&
  {Begelman}}{{Salvesen} et~al.}{2016}]{SSAB16}
{Salvesen} G.,  {Simon} J.~B.,  {Armitage} P.~J.,   {Begelman} M.~C.,  2016,
  \mn@doi [\mnras] {10.1093/mnras/stw029}, \href
  {https://ui.adsabs.harvard.edu/abs/2016MNRAS.457..857S} {457, 857}

\bibitem[\protect\citeauthoryear{{Siess}, {Forestini}  \& {Bertout}}{{Siess}
  et~al.}{1997}]{SFB97}
{Siess} L.,  {Forestini} M.,   {Bertout} C.,  1997, \aap, \href
  {https://ui.adsabs.harvard.edu/abs/1997A&A...326.1001S} {326, 1001}

\bibitem[\protect\citeauthoryear{{Suzuki}, {Muto}  \& {Inutsuka}}{{Suzuki}
  et~al.}{2010}]{SMI10}
{Suzuki} T.~K.,  {Muto} T.,   {Inutsuka} S.-i.,  2010, \mn@doi [\apj]
  {10.1088/0004-637X/718/2/1289}, \href
  {https://ui.adsabs.harvard.edu/abs/2010ApJ...718.1289S} {718, 1289}

\bibitem[\protect\citeauthoryear{{Terquem} \& {Papaloizou}}{{Terquem} \&
  {Papaloizou}}{2007}]{TP07}
{Terquem} C.,  {Papaloizou} J. C.~B.,  2007, \mn@doi [\apj] {10.1086/509497},
  \href {https://ui.adsabs.harvard.edu/abs/2007ApJ...654.1110T} {654, 1110}

\bibitem[\protect\citeauthoryear{{Tsang}}{{Tsang}}{2011}]{Tsang11}
{Tsang} D.,  2011, \mn@doi [\apj] {10.1088/0004-637X/741/2/109}, \href
  {https://ui.adsabs.harvard.edu/abs/2011ApJ...741..109T} {741, 109}

\bibitem[\protect\citeauthoryear{{Udry} et~al.,}{{Udry} et~al.}{2007}]{UBD07}
{Udry} S.,  et~al., 2007, \mn@doi [\aap] {10.1051/0004-6361:20077612}, \href
  {https://ui.adsabs.harvard.edu/abs/2007A&A...469L..43U} {469, L43}

\bibitem[\protect\citeauthoryear{{Ueda}, {Okuzumi}  \& {Flock}}{{Ueda}
  et~al.}{2017}]{UOF17}
{Ueda} T.,  {Okuzumi} S.,   {Flock} M.,  2017, \mn@doi [\apj]
  {10.3847/1538-4357/aa74b5}, \href
  {https://ui.adsabs.harvard.edu/abs/2017ApJ...843...49U} {843, 49}

\bibitem[\protect\citeauthoryear{{Ward}}{{Ward}}{1997}]{W97}
{Ward} W.~R.,  1997, \mn@doi [\icarus] {10.1006/icar.1996.5647}, \href
  {https://ui.adsabs.harvard.edu/abs/1997Icar..126..261W} {126, 261}

\bibitem[\protect\citeauthoryear{{Weidenschilling} \&
  {Marzari}}{{Weidenschilling} \& {Marzari}}{1996}]{WM96}
{Weidenschilling} S.~J.,  {Marzari} F.,  1996, \mn@doi [\nat]
  {10.1038/384619a0}, \href
  {https://ui.adsabs.harvard.edu/abs/1996Natur.384..619W} {384, 619}

\bibitem[\protect\citeauthoryear{{Youdin}}{{Youdin}}{2011}]{You11}
{Youdin} A.~N.,  2011, \mn@doi [\apj] {10.1088/0004-637X/742/1/38}, \href
  {https://ui.adsabs.harvard.edu/abs/2011ApJ...742...38Y} {742, 38}

\bibitem[\protect\citeauthoryear{{Zeng} et~al.,}{{Zeng} et~al.}{2019}]{Zeng19}
{Zeng} L.,  et~al., 2019, \mn@doi [Proceedings of the National Academy of
  Science] {10.1073/pnas.1812905116}, \href
  {https://ui.adsabs.harvard.edu/abs/2019PNAS..116.9723Z} {116, 9723}

\makeatother
\end{thebibliography}

\appendix

\section{Disk Structure}

\subsection{Magnetic Field Diffusion in Disk}\label{Ap:mag_diffuse}
As mentioned in section~\ref{sss:trans_zone_alpha}, $B_{nor}$ is calculated by requiring the total flux going through the disk (up to 2AU) to be the same across all n value given the same stellar $B$-field. For convenience, we chose to normalize to the dipole case ($n=3$) which gives the following normalized magnetic field for all $n$ value except $n=2$:
\begin{equation}
    B_{nor} = B_0R_*^{3-n}(n-2)\frac{1/R_{out}-1/R_{in}}{R_{out}^{2-n}-R_{in}^{2-n}}
\end{equation}\label{eq:Cn_not_n2}

Combining this normalization scheme with equation~\ref{eq:B_loc_main}, it can be inferred that decreasing the power-law index ($n$) leads to a decrease in local magnetic field interior to 0.1~AU and an increase exterior to 0.1~AU. This change in local magnetic field is the underlining cause of the change in surface density described in section~\ref{ss:M_disk_structure}.

\subsection{Disk Surface Density and Temperature Derivation}\label{ap:ST_derivation}

\subsubsection{Passive Heating}\label{apss:passive_heating}

We adopted the passive heating model from \cite{UOF17} where the authors split the disk into 4 different regions. We found that we can write down an approximate expression as the following:

\begin{equation}
    T_p =
\left\{
	\begin{array}{ll}
		\left(\frac{R_*}{2R}\right)^{\frac{1}{2}}T_*  & \mbox{if } R \leq R_{AB} \\
		\left(\frac{\arctan\left(\frac{R_{CD} - R}{h}\right)+\arctan\left(\frac{R_{BC}-R}{w}\right)+\pi}{2\pi}\right)^{\frac{1}{4}}T_{ev} & \mbox{if } R > R_{AB}
	\end{array}
\right.
\label{eq:Ueda_T_full}
\end{equation}

where $T_{ev}$ is the evaporative temperature set by calcium aluminum inclusions. The radii $R_{AB}$, $R_{BC}$ and $R_{CD}$ are the locations marking the transitions between regions in \cite{UOF17} and $R_{AB} < R_{BC} <R_{CD}$. 

The width variables at $R_{BC}$ and $R_{CD}$ are given by $w$ and $h$ respectively. Constants and equations for variables used in equation~\ref{eq:Ueda_T_full} are given as the following:

\begin{equation}
    \begin{split}
        T_{ev} &= 2000K\left(\frac{\rho}{1g\cdot cm^{-3}}\right)^{0.0195} \\
        \textit{where} \\
        \rho &= \frac{\Sigma}{\sqrt{2\pi}}
    \end{split}
    \label{eq:T_ev}
\end{equation}

\begin{equation}
    R_{AB} = \frac{1}{2}\left(\frac{T_*}{T_{ev}}\right)^2R_*
    \label{eq:R_AB}
\end{equation}

\begin{equation}
    R_{BC} = \frac{1}{2}\left(\frac{\kappa_d(T_*)}{\kappa_d(T_ev)}\right)^{\frac{1}{2}}\left(\frac{T_*}{T_{ev}}\right)^2R_*
    \label{eq:R_BC}
\end{equation}

\begin{equation}
\begin{split}
    R_{CD} = & R_{BC}\times\sqrt{1+\Gamma} \\
    \Gamma = & 3.1\left(\frac{R_{BC}}{0.46au}\right)^{-12/7}\left(\frac{T_*}{10^4K}\right)^{32/7}\left(\frac{M_*}{2.5M_\odot}\right)^{-1/2} \\
    & \times\left(\frac{R_*}{2.5R_\odot}\right)^{16/7}
\end{split} \label{eq:R_CD}
\end{equation}

\begin{equation}
    h = 0.05R_{CD}
    \label{eq:Ueda_h}
\end{equation}

\begin{equation}
    w = H(R_{BC})
    \label{eq:passive_heating_w}
\end{equation}

The main difference we have compared to \cite{UOF17} is that of equation~\ref{eq:passive_heating_w}. We assumed disk structures should be smoothed out over at least 1 scale height at a given location. 

The local passive heating term is then:

\begin{equation}
    PH(R) = 4\sigma T_{p}^4
    \label{eq:passive_heating_term}
\end{equation}

\section{Normalization Used in Torque Maps}\label{ap:torque_nor}
As the torque values span a few orders of magnitude in the mass and radius range we considered, we employed a normalization scheme in order to show the features clearly in figure~\ref{ss:F_migmap} and figure~\ref{fig:migration_map_100_250}. We normalized all torque values to the absolute value of the torque of a 1 $M_{\odot}$ planet at 1AU and then we further removed the mass dependency in the coefficient of the torques by dividing with their respective planet masses squared. In practise, this is done by:

\begin{equation}
    \Gamma_{nor} = \frac{1}{{M_p}^2}\frac{\Gamma(R,M)}{\lvert\Gamma(1AU,1M_{\oplus})\rvert}
\end{equation}
, where $\Gamma_{nor}$ is the normalized torque shown in the figures, $M_p$ is the planet mass in unit of Earth masses and $\Gamma(R,M)$ is the torque calculated based on the torque prescription (see section~\ref{ss:Type1_prescription}) for a planet of mass $M$ at a radial location of $R$.

\section{Decoupling under shallow magnetic profile}\label{AP:decoupling_low_n}
 We find that the decoupling behavior deviates from the generic case when the power-law index of the magnetic field profile (n) drops below 2.65. We further divide the shallow magnetic profile behavior into two types, the incomplete deactivation track and the fall-through track.

\subsection{Incomplete Deactivation Track}\label{apss:dis_trac}
\begin{figure}
    \centering
    \includegraphics[width=\linewidth]{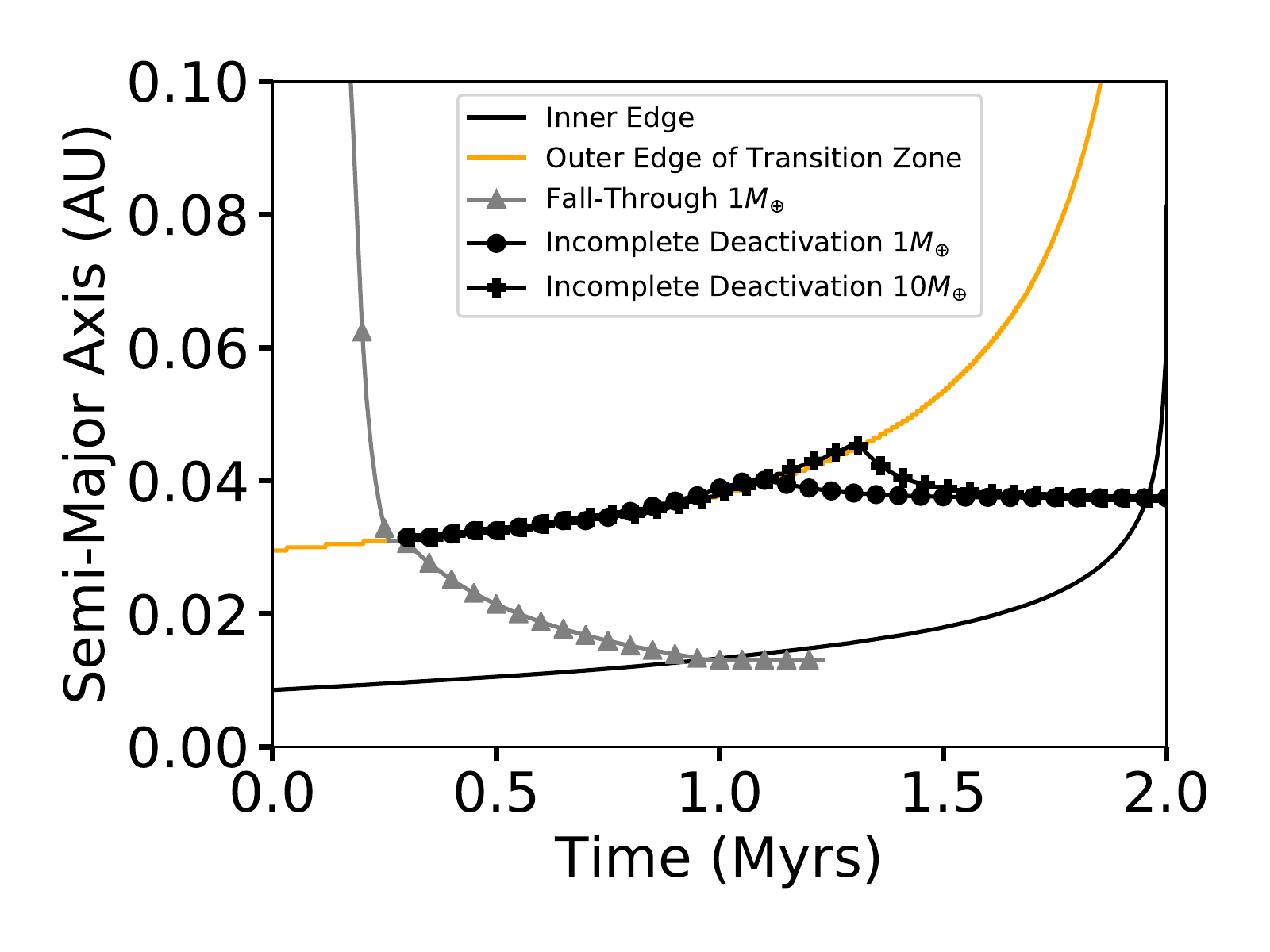}
    \caption{Showing examples of the incomplete-deactivation and fall-through migration behavior. 
    The incomplete-deactivation behavior is scale down to the current location by multiplying by a factor of $\sim0.8$.
    In the incomplete deactivation case, planets initially behave  identically to the generic track for the most part up to the decoupling phase.
    During decoupling, instead of gradually migrating outwards and arriving at their final location asymptotically, planets are observed to migrate further inward towards the host star and arrive at a closer location that is largely independent to planet mass. The initial inward migration for both 1 and 10$M_{\oplus}$ are omitted for clarity.
    In the Late-arrival case, planets migrate inward as well but the transition zone cannot completely stop the migrating planet. Instead, planets merely slow their migration rates and eventually migrate through the inner edge of the disk, into the cavity.}
    \label{fig:migration_track2}
\end{figure}

\begin{figure}
    \centering
    \includegraphics[width=\linewidth]{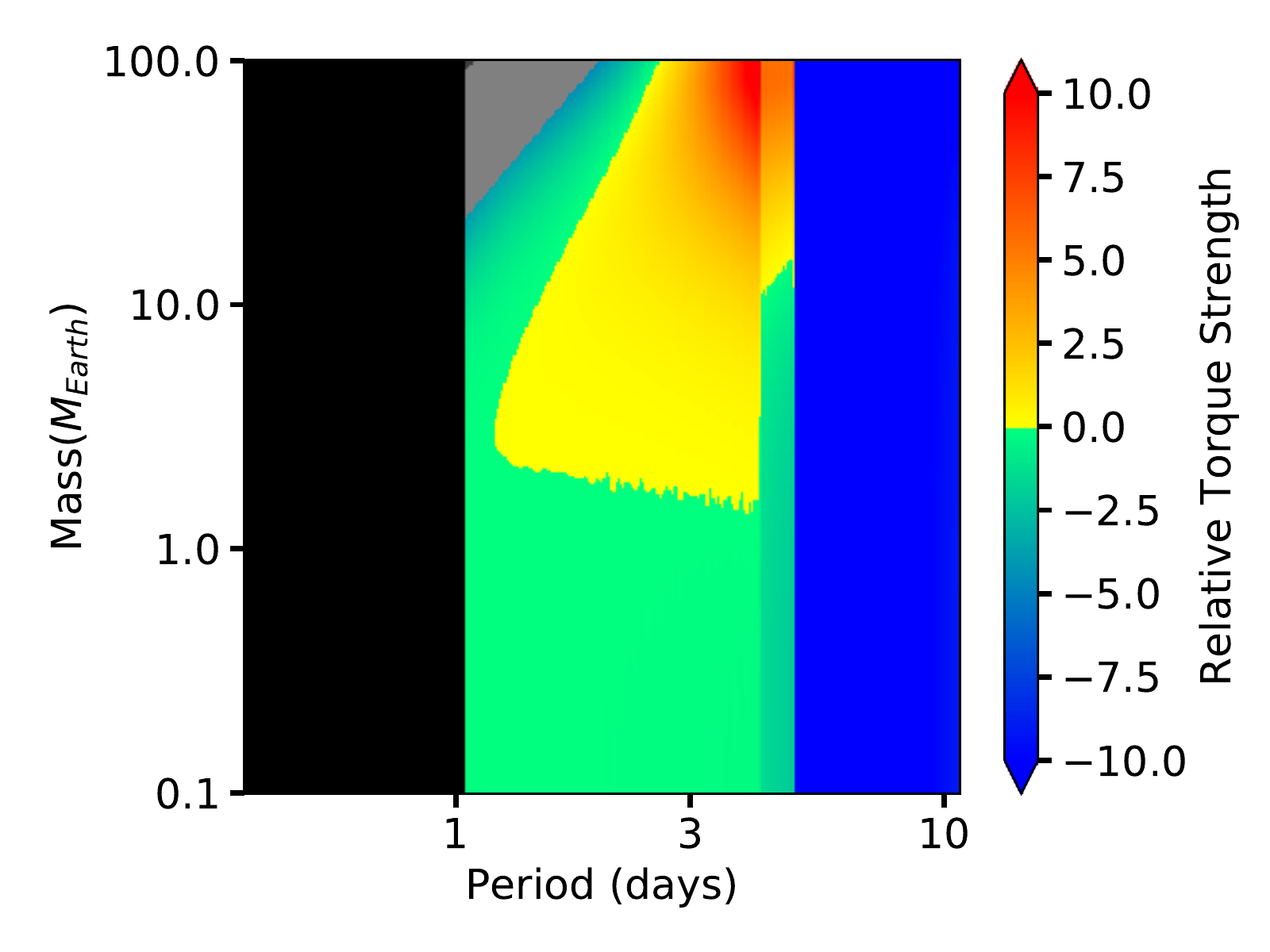}
    \caption{Migration map for the disk with $B = 100$ and $n = 2.50$ at 1.15Myrs --- shortly after the decoupling process starts for the incomplete deactivation case. The outward migration area (red and yellow) is partially deactivated for planets less than 10 $M_{\oplus}$. This deactivation begins at the lower mass end and moves up as the accretion rate decreases. As part of the transition zone no longer support outward migration after this deactivation, planets migrate inward towards the next stable location, the transition between temperature zone i and ii, and become trapped there until the disk disperse regardless of their masses.}
    \label{fig:migration_map_100_250}
\end{figure}

The incomplete deactivation track is seen in systems where the torque reversal associated with the transition zone becomes partially deactivated.
This track is identical to the generic track for the most part up to the decoupling phase.
During decoupling, instead of gradually migrating outwards and arriving at their final location asymptotically, planets are observed to migrate further inward towards the host star and arrive at a much closer location that is largely independent to planet mass.
This can be seen in Figure~\ref{fig:migration_track2} where we shown the migration track for this particular scenario for a 1 and 10 $M_{\oplus}$ planet.
The two arrived at virtually the same final semi-major axis, contrary to the case in the generic track.
The cause of this behavior can be seen in the snapshot migration map in Figure~\ref{fig:migration_map_100_250} which shows the relative strength of the torque during the decoupling phase shown in Figure~\ref{fig:migration_track2}.
Using the same color code as in Figure~\ref{fig:disk_map}, we can see the outward migration area is partially deactivated for planets less than 10 $M_{\oplus}$.
This deactivation begins at the lower mass end,which happens before the figure shown, and moves up as the accretion rate decreases.
The cause of this deactivation of the outward torque is related to a combination between the much shallower power-law index in the surface density and the decreasing accretion rate over time.
As pointed out in section~\ref{ss:M_disk_structure}, smaller power-law index(n) leads to a shallower surface density power-law in the transition zone.
In turn, the corotation torques that are required for planet trapping are weakened.
This effect is more obvious on the lower mass planets due to their smaller horseshoe orbits which are the sources of their outward torque.
The combined effect leads the lower mass planets to be released from the trap earlier than the higher mass planets, which results in slightly different migration tracks between the 1 and 10 $M_{\oplus}$ case.
As part of the transition zone no longer support outward migration after this deactivation, planets migrate inward towards the next stable location, the transition between temperature zone i and ii, and become trapped there until the disk disperse regardless of their masses.
The disrupted type is only seen in the $B=100G$, $n=2.50$ simulations.
We postulate that the incomplete decoupling type is the transition between the generic type and the fall-through type that we will discuss next.

\subsection{Fall-through Track}\label{apss:fall_trac}

Finally, we also see a set of evolutions which we call the fall-through case.
This can be seen as a variant of the late-arrival track.
Both of them share the similarity that the planets migrate inward through the majority of the disk and do not couple to the planet trap at the outer edge of the transition zone.
However, in the fall-through type, the planets arrive at the transition zone much earlier and still have higher level of interaction with the transition zone compared to the late-arrival case.
This behavior is shown in fig~\ref{fig:migration_track2} where a 1 $M_{\oplus}$ planet is shown to migrate through the majority of the disk.
However, once it reached the outer edge of the transition zone, instead of coupling to the would-be planet trap at the outer edge transition zone, the effect of the change in surface density is to merely slow the rate of inward evolution, and the planet only stops at the inner truncation of the disk.
This case is seen when the influence of the stellar field is weakest -- either because $B$ or $n$ is small.
It is also worth noting that we observe similar behavior where the planets migrate inward through the entire disk unimpeded if the magnetic field is set to 0.

\dotfill

\end{document}